\documentclass[longauth]{aa}
\usepackage{caption}

\usepackage{txfonts}
\usepackage{graphicx}
\usepackage[switch,columnwise]{lineno}
\makeatletter
\let\linenumbers\relax

\makeatother
\usepackage[dvipsnames]{xcolor}
\usepackage[unicode=true,psdextra]{hyperref}
\begin{document}

\title{Optical spectroscopy of blazars for the Cherenkov Telescope Array Observatory- IV
\thanks{ Based on observations collected at the European Organisation for Astronomical Research in the Southern Hemisphere, Chile, under programs P108.22CJ001 P109.238K.001 P111.24MZ.002. The raw FITS data files are available in the ESO archive. Some of the data presented herein were obtained at the W. M. Keck Observatory, which is operated as a scientific partnership among the California Institute of Technology, the University of California and the National Aeronautics and Space Administration. The Observatory was made possible by the generous financial support of the W. M. Keck Foundation. Based on observations made with the Southern African Large Telescope (SALT) under program 2021-1-MLT-008 (PI E. Kasai)}.}

   \author{B. Rajput   \inst{1}
           \and 
           P. Goldoni \inst{2}
           \and 
           W. Max-Moerbeck\inst{1}
           \and 
           E. Kasai \inst{3} 
           \and
           D. A. Williams \inst{4}
           \and
           C. Boisson \inst{5}
           \and 
           S. Pita \inst{6}
           \and
           M. Backes \inst{3,7} 
           \and
           U. Barres de Almeida \inst{8}
           \and
           J. Becerra González \inst{9,10}
           \and
           G. Cotter \inst{11}
           \and
           F. D'Ammando \inst{12}
           \and
           V. Fallah Ramazani  \inst{13}
           \and
           B. Hnatyk \inst{14}
           \and
           O. Hervet \inst{4}
           \and
           E. Lindfors \inst{13}
           \and
           D. Mukhi-Nilo \inst{15}
           \and
           M. Nikołajuk \inst{16}
           \and
           M. Splettstoesser \inst{4}
           \and
           B. Van Soelen \inst{17}
           }

   \institute{Departamento de Astronomi{\'a}, Universidad de Chile, Camino El Observatorio 1515, Las Condes, Santiago, Chile\\
              \email{bhoomikarjpt2@gmail.com}
         \and Universit\'{e} Paris Cit\'{e}, CNRS, CEA, Astroparticule et Cosmologie, F-75013 Paris, France
         \and Department of Physics, Chemistry \& Material Science, University of Namibia, Private Bag 13301, Windhoek, Namibia
         \and Santa Cruz Institute for Particle Physics and Department of Physics, University of California, Santa Cruz, CA 95064, USA
         \and Laboratoire Univers et Th\'{e}ories, Observatoire de Paris, Universit\'{e} PSL, Universit\'{e} Paris Cit\'{e}, CNRS, F-92190 Meudon, France
         \and Universit\'{e} Paris Cit\'{e}, CNRS, Astroparticule et Cosmologie, F-75013 Paris, France
         \and Centre for Space Research, North-West University, Potchefstroom 2520, South Africa
         \and Centro Brasileiro de Pesquisas Fisicas (CBPF), Rua Dr. Xavier Sigaud 150 - Urca, Rio de Janeiro 22290-180, Brazil
         \and Universidad de La Laguna (ULL), Departamento de Astrof\'isica, E-38206 La Laguna, Tenerife, Spain
         \and Instituto de Astrof\'isica de Canarias (IAC), E-38200 La Laguna, Tenerife, Spain
         \and Oxford Astrophysics, University of Oxford, Denys Wilkinson Building, Keble Road, Oxford, OX1 3RH, United Kingdom
         \and INAF - Istituto di Radioastronomia, Via Gobetti 101, I-40129 Bologna, Italy 
         \and Finnish Centre for Astronomy with ESO, FINCA, University of Turku, Turku, FI-20014 Finland
         \and Astronomical Observatory of Taras Shevchenko National University of Kyiv, 3 Observatorna Street, Kyiv, 04053, Ukraine
         \and Instituto de Astrof\'{i}sica, Facultad de Fis\'{i}ca, Pontificia Universidad Cat\'{o}lica de Chile, Av. Vicuña Mackenna 4860, Macul, Santiago, Chile 
         \and University of Białystok, Faculty of Physics, ul. K. Ciołkowskiego 1L, 15-245 Białystok, Poland
         \and Department of Physics, University of the Free State, Bloemfontein 9300, South Africa}

   \date{Received / Accepted}

\linenumbers
 
  \abstract
   { Blazars, comprising BL Lacertae objects (BL Lacs) and flat-spectrum radio quasars (FSRQs), are the most luminous extragalactic sources that dominate the $\gamma$-ray sky. They account for approximately 56\% of the sources listed in the recent {\em Fermi}-LAT (Large Area Telescope) catalog (4FGL$-$DR4). The optical and UV spectra of BL Lacs are nearly featureless, making it difficult to precisely determine their redshifts. Consequently, nearly half of the $\gamma$-ray BL Lacs lack reliable redshift measurements. This poses a significant challenge, since redshift is crucial for studying the cosmic evolution of the blazar population and for understanding their intrinsic emission mechanisms. Additionally, it is vital for $\gamma$-ray propagation studies, such as indirect evidence of extragalactic background light (EBL), placing constraints on the intergalactic magnetic field (IGMF), and searches for Lorentz invariance violation (LIV) and axion-like particles (ALPs).}
   { This paper is the fourth in a series dedicated to determining the redshift of a sample of blazars identified as key targets for future observations with the Cherenkov Telescope Array Observatory (CTAO). The precise determination of the redshifts of these objects plays a crucial role in planning future CTAO observations.}
   { We carried out Monte Carlo simulations to identify potential $\gamma$-ray blazars with hard spectra detected by the {\em Fermi}-LAT telescope that currently lack redshift measurements. These simulations selected the blazars that are anticipated to be detectable by the CTAO within 30 hours or less of exposure assuming an average flux state. In this fourth paper, we report the results of detailed spectroscopic observations of 29 blazars using the ESO/VLT, Keck II, and SALT telescopes. Our analysis involved a thorough search for spectral lines in the spectra of each blazar, and when features of the host galaxy were identified, we modeled its properties. Moreover, we compared the magnitudes of the targets during the observations to their long-term light curves.}
   {In the sample studied, 9 of 29 sources were observed with a high signal-to-noise ratio (S/N $>$ 100), while the remaining 20 were observed with a moderate or low S/N. We successfully determined firm redshifts for 12 blazars, ranging from 0.1636 to 1.1427, and identified two lower limit redshifts at $z > 1.0196$ and $z > 1.4454$. The remaining 15 BL Lac objects exhibited featureless spectra under the observed S/N ratio.}
   {}

   \keywords{galaxies: active - BL Lacertae objects: general - gamma-rays: galaxies - galaxies: distances and redshifts
               }
               \maketitle
%

\section{Introduction}

  Blazars are a distinctive class of jetted active galactic nuclei (AGN), with powerful relativistic jets directed along our line of sight \citep[see e.g.][]{1993ARA&A..31..473A, 1995PASP..107..803U, 2017A&ARv..25....2P}. As a result, these jets appear to be strongly Doppler boosted, giving blazars their unique properties \citep[see e.g.][]{1979ApJ...232...34B}. Their unique characteristics include strongly-beamed non-thermal emission that spans the entire electromagnetic spectrum, from low-energy radio to high-energy $\gamma$-rays \citep[see e.g.][]{2019NewAR..8701541H, 2019ARA&A..57..467B}, and rapid, high-amplitude flux variations throughout the \-observable spectrum, occurring over timescales ranging from minutes to years \citep[see e.g.][]{Wag95, 1997ARA&A..35..445U,2007ApJ...664L..71A, 2011ApJS..194...29R, Falo14, 2016ApJ...824L..20A, 2019ApJ...877...39M}. Furthermore, blazar jets show detectable polarization in optical \citep[see e.g.][]{Angel80,1990A&AS...83..183M,Angelakis16}, radio \citep{Lis11} and X-rays \citep{2022ApJ...938L...7D, 2022Natur.611..677L, 2023NatAs...7.1245D}, with variable polarization \citep[see e.g.][]{2010Natur.463..919A, 2017ApJS..232....7F}. Some also show the ejection of superluminal radio blobs \citep[see e.g.][]{VerCo94}.

   Since 2008, the Large Area Telescope (LAT) on board the {\em Fermi Gamma Ray Space Telescope} (referred to hereafter as {\em Fermi}-LAT\footnote{\url{https://fermi.gsfc.nasa.gov/}}) \citep{LAT} has opened a new era of the study of blazars in the realm of high-energy $\gamma-$ray astronomy (100 MeV $<$ $E$ $<$ 300 GeV). The {\em Fermi}-LAT has identified 7194 $\gamma-$ray sources, of which 3933 are blazars (4FGL-DR4; \citealt{2023arXiv230712546B}). Of these 3933 blazars, 1490 are BL Lacs, 820 are flat-spectrum radio quasars (FSRQs), and 1623 are blazar candidates of uncertain type (BCUs).  Notably, 90\% of the BCUs lack redshift estimates, and 36\% of the BL Lac objects also do not have redshift measurements. In the past two decades, very high-energy (VHE $>$ 100 GeV) $\gamma-$ray astronomy has also advanced greatly, largely thanks to the efforts of the current generation of imaging atmospheric Cherenkov telescopes (IACTs): VERITAS\footnote{\url{https://veritas.sao.arizona.edu/}}, H.E.S.S.\footnote{\url{https://www.mpi-hd.mpg.de/HESS/}}, and  MAGIC\footnote{\url{https://magic.mpp.mpg.de/}} \citep[see e.g.][]{VERITAS, HESS, MAGIC2,MAGIC1}. Among the 309 sources recognized as TeV sources \citep[TeVCAT\footnote{\url{http://tevcat.uchicago.edu/}};][]{Wakely08}, 
    99 are extragalactic, 85\% of which are classified as blazars, as of February 2025. VHE emission has been detected in both types of blazars: FSRQs and BL Lacs, where BL Lacs account for the majority of VHE-detected blazars. However, the redshifts of 12 out of 70 VHE BL Lacs are still unknown, whereas all 10 FSRQs have their redshifts measured. 

    Determining the redshift of BL Lac objects remains challenging \citep[see e.g.][]{2013ApJ...764..135S, Pai17a, Pena20}. This difficulty arises because their optical spectra usually show weak emission lines with equivalent width (EW) $<$ 5\AA, if any at all, in contrast to FSRQs, which display prominent emission lines (EW $>$ 5\AA) \citep[e.g.][]{1991ApJS...76..813S}. The detection of spectral features is complicated by the non-thermal variable jet emission, which follows a featureless power-law distribution and which often overwhelms the emission and absorption lines from the host galaxy. 
    To detect such weak emission from BL Lacs, their spectra need to have a high S/N ratio. Despite numerous efforts that have been made using X-ray and radio-selected blazars, which were limited to bright sources \citep[BZCAT\footnote{\url{https://www.ssdc.asi.it/bzcat/}};][]{bzcat15}, the redshift sample for BL Lac objects is still inadequate. 
   Indeed the redshift completeness of the best BL Lac samples is about 40-50 $\%$ \citep[see e.g.][]{2013ApJ...764..135S}. Given this situation, other methods can be used to constrain the redshift such as deep imaging of the host galaxy \citep{2003A&A...400...95N, 2024A&A...691A.154N} and association of the BL Lac object to a coincident galaxy group \citep{2024MNRAS.531.5084K}. The limited redshift data for BL Lac objects also impacts the study of their cosmic evolution \citep[see e.g.][]{Aje14}. This issue worsens at VHE because of the very limited number of sources detected at higher redshifts due to absorption by the EBL.
   
    In the next few years, the Cherenkov Telescope Array Observatory (CTAO\footnote{\url{https://www.cta-observatory.org/}}) will become operational, routinely achieving a lower energy threshold for VHE $\gamma$-ray detections down to several tens of GeV. Moreover, it will have a flux sensitivity enhanced by approximately a factor of 10 in the TeV energy range compared to existing facilities. 
    The wide energy range of the CTAO, spanning 20 GeV to 300 TeV, will improve the detectability of many AGNs and enable more comprehensive population studies. The CTAO will enable the detection of $\gamma$-rays from blazars whose VHE spectra undergo significant distortion due to interactions with EBL. This will place critical constraints on both the intensity of the EBL and the intrinsic spectra of the blazars \citep[see e.g.][]{2013APh....43..112D, Bit15}. The EBL holds valuable information about the formation and evolution of stars and galaxies over cosmic time, and blazar studies provide an indirect method to probe these. The electron-positron pairs produced by $\gamma$-ray interactions with the EBL are sensitive to the IGMF, which remains poorly understood. These pairs can therefore be used to constrain the properties of the IGMF  \citep[see e.g.][]{1994ApJ...423L...5A, 2013A&ARv..21...62D,2019MNRAS.489.3836A}. The propagation of VHE $\gamma$-ray radiation can also be used to address major questions in cosmology and fundamental physics, including the potential existence of ALPs \citep[see e.g.][]{2007PhRvD..76b3001M, 2013PhRvD..88j2003A}, searches for LIV \citep{1999ApJ...518L..21K}, and offering an independent method to determine the Hubble constant, $H_0$ \citep{1994ApJ...423L...1S}. BL Lac objects play a crucial role in determining the density of EBL \citep[see e.g.][]{Aje14, Bit15}, especially those with redshifts greater than 0.3 \citep[see][]{2019scta.book.....C}. However, an accurate determination of the redshift for these sources is key to the success of such studies. Therefore, it is essential to measure the redshifts for a significant number of blazars detected by {\em Fermi}-LAT, as these blazars, based on their extrapolated LAT spectra, are excellent candidates for CTAO observations. Determining their redshifts will help in selecting the most suitable ones for these observations.

   This work is the fourth paper in a series that aims to determine spectroscopically the redshifts of blazar samples that the CTAO is most likely to detect. In three previous papers, \citet[hereafter Paper~I]{2021A&A...650A.106G}, \citet[hereafter Paper~II]{2023MNRAS.518.2675K} and \citet[hereafter Paper~III]{2024A&A...683A.222D}, the firm redshifts of 37 blazars, within the range from 0.0838 to 0.8125, along with 2 tentative redshifts and 6 lower limits were established. In this paper, we provide comprehensive results for 29 new targets. 
   
   The paper is structured as follows: Sections \ref{obs plan} and \ref{Obs and data} provide details of the sample and observing strategy, and of the observations and data reduction, respectively. In Sections \ref{zmeas} and \ref{sources}, we detail the analysis and results, and in Section \ref{lcurves} we discuss the flux of the targets during our observations, using public photometric light curves to place the observed flux in the context of the sources' activity. Finally, Section \ref{Conc} contains the discussion and conclusions.

\section{Sample and observing strategy}\label{obs plan}    
We aim to determine the spectroscopic redshifts, or establish lower limits, for BL Lac objects and BCUs. These sources are selected from the Third \emph{Fermi}-LAT Catalog of High-Energy Sources \citep[3FHL;][]{Fer3FHL17}, that spans the energy range from 10 GeV to 2 TeV, and where more than 50\% of the sources lack redshift measurements. The sample selection process is thoroughly explained in Paper I, Paper II, and Paper III. Briefly, the procedure involved extrapolating the {\em Fermi}-LAT spectra into the TeV domain and performing Monte Carlo simulations using the Gammapy\footnote{\url{https://gammapy.org}} software \citep{gammapy:2023} in conjunction with the publicly accessible CTAO performance files.\footnote{\url{https://www.cta-observatory.org/wp-content/uploads/2019/04/CTA-Performance-prod3b-v2-FITS.tar.gz}}$^{, }$\footnote{\url{https://zenodo.org/record/5163273\#.Yg9-yPVBzPZ}} EBL absorption was applied using the model by \citet{Dom11}. For 3FHL sources with no redshift available, we assumed a value of $z = 0.3$ similar to the median values in \citet{2013ApJ...764..135S} and \citet{Pena20}. Sources detectable with the CTAO in less than 30 hours were selected for the sample.

The details of the objects analyzed in this work are provided in Table \ref{tabobs1}. All the sources, except SUMSS\,J052542$-$601341, 4C\,+29.48, 1RXS\,J171405.2$-$202747, NVSS\,J182338$-$345412 and NVSS\,J192502+28154 are identified as BL Lac objects in the 4FGL-DR4 catalog. These five sources are classified as BCUs in the 4FGL-DR4 catalog and high-synchrotron peaked BL Lacs (HBLs) in the 3HSP blazar catalog \citep{3HSP}.

BL Lacs are mainly hosted by luminous elliptical galaxies \citep{Urr00}. To achieve our objective, we focused on detecting stellar absorption features in the host galaxies of BL Lac objects, including the CaHK doublet, Mgb, and NaID. Although emission lines such as [OII], H$\beta$, [OIII], H$\alpha$, and [NII] are rarely detected in the BL Lac spectra, we also incorporated them into our investigations. A more complete line list can be found in Paper~II. These objects typically exhibit weak lines with an $\lvert EW \rvert$ of less than 5 \AA. Detecting such faint lines requires a spectral resolution of several hundred (ideally near 1000) and a S/N of around 100 per pixel. This combination of conditions has been shown to be effective in previous studies, as demonstrated in Paper I, Paper II, Paper III, and in \citet{Pit14}. If the chosen instrument cannot provide spectra with both of these properties, we configure it to obtain at least one of them. Additionally, observing these sources during periods of low optical activity could further improve the S/N by reducing the non-thermal foreground from the AGN (see Paper~III and Section 6).

\section{Observations and data reduction}\label{Obs and data}   
We observed 29 blazars between April 3, 2021, and June 20, 2023, using three instruments mounted on separate telescopes: the Keck/ESI\footnote{\raggedright Echellette Spectrograph and Imager (ESI) on the Keck II telescope, \url{https://www.keckobservatory.org/about/telescopes-instrumentation}}~\citep{Shei12}, SALT/RSS\footnote{Robert Stobie Spectrograph (RSS) on the Southern African Large Telescope (SALT), \url{www.salt.ac.za/telescope}}~\citep{Burgh03}, and VLT/FORS\footnote{FOcal Reducer and low dispersion Spectrograph (FORS) on the Very Large Telescope (VLT), \url{https://www.eso.org/sci/facilities/paranal/instruments/fors.html}}~\citep{Appenzeller1998} for a total observing time of approximately 36 hours. The details of the observations are provided in Table \ref{tabobs1}. 

 Details of data reduction, flux calibration, telluric corrections, and spectral dereddening are presented in Paper~I, Paper~II, and Paper~III and we have followed the same approach in this study.
The comprehensive description of the Keck/ESI and SALT/RSS instruments is given in Paper I, and a description of VLT/FORS is provided in Paper III. The main parameters of the instruments and the configuration employed are summarized in Table \ref{tab_spgraphtech}.

 \begin{sidewaystable*}
\small
\caption{ \label{tabobs1} The list of observed sources and the observation parameters for the sample consisting of 29 sources, which are discussed in detail in Section \ref{sources}. }
\centering
\begin{tabular}{lcccclcclll}
\hline\hline
3FHL Name &   4FGL Name & Source Name   & R.A. & Dec.  &   Telescope/ &   Slit            & Start Time & Exp.  & Airm. & Seeing      \\
                   &                         &         &  (J2000)      &    (J2000)        &    Instrument &  (\arcsec)  &  UTC         &  (sec)   &         & (\arcsec)      \\  
 (1)             & (2)                    &             (3)        &  (4) & (5)  &      (6)            &     (7)          &      (8)       &   (9)      &(10)    & (11)           \\ 
\hline
3FHL\,J0323.6$-$0109  & 4FGL\,J0323.7$-$0111  & 1RXS\,J032342.6$-$011131$^{\dagger}$ & 03 23 43.6 & $-$01 11 46.21 & VLT/FORS & 1.3 & 2022-09-23 07:18:38 & 1600 &1.10 & 0.7 \\
3FHL\,J0338.5+1302 &  4FGL\,J0338.5+1302 &  RX\,J0338.4+1302  &  03 38 29.3 & +13 02 15 &  Keck/ESI &   1.0 &  2021-10-15 11:06:53  &  7200 & 1.03  & 1.1 \\
3FHL\,J0525.6$-$6013 & 4FGL\,J0525.6$-$6013 &  SUMSS\,J052542$-$601341 &  05 25 42.4 & -60 13 40 &  SALT/RSS  & 2.0 &  2021-10-07 23:55:47  & 2400 &  1.26 &  1.3 \\ 
& & & & & & & 2021-10-08 23:44:56  & 1990 & 1.27 &  1.3 \\
3FHL\,J0540.5+5823 &  4FGL\,J0540.5+5823  & GB6\,J0540+5823$^{\dagger}$  & 05 40 30.0 & +58 23 40 & Keck/ESI &  1.0 &  2021-10-15 13:30:38  & 5900 &  1.29 &  1.0\\
3FHL\,J0622.4$-$2606 & 4FGL\,J0622.3$-$2605 & PMN\,J0622$-$2605 & 06 22 23.7 & $-$26 06 28 & VLT/FORS & 1.3 & 2022-02-06 03:31:47 & 1730 & 1.10 & 0.7\\ 
3FHL\,J0702.6$-$1950 & 4FGL\,J0702.7$-$1951 & TXS\,0700$-$197$^{\dagger}$ & 07 02 42.9 & $-$19 51 22 & VLT/FORS & 1.3 & 2022-02-06 02:00:23 & 2595 & 1.01 & 0.7\\
3FHL\,J0804.0$-$3629 & 4FGL\,J0804.0$-$3629 & NVSS\,J080405$-$362919 & 08 04 05.3 & $-$36 29 19 & VLT/FORS & 1.3 & 2022-03-10 00:39:07 & 2430 & 1.03 & 1.1\\
& & & & & & & 2022-05-02 00:05:33 & 2430 & 1.18 & 0.8\\ 
3FHL\,J0819.4$-$0756 & 4FGL\,J0819.4$-$0756 & RX\,J0819.2$-$0756$^{\dagger}$ & 08 19 17.6 & $-$07 56 26 & VLT/FORS & 1.3 & 2022-02-25 02:10:22 & 5190 & 1.09 & 0.7\\
3FHL\,J0841.3$-$3554 & 4FGL\,J0841.3$-$3554 & NVSS\,J084121$-$355506$^{\dagger}$ & 08 41 21.6 & $-$35 55 05 & VLT/FORS & 1.3 & 2022-03-04 01:01:02 & 2595 & 1.06 & 0.6\\
3FHL\,J0947.2$-$2541 & 4FGL\,J0947.1$-$2541 & 1RXS\,J094709.2$-$254056 & 09 47 09.5 & $-$25 41 00  & SALT/RSS & 2.0   & 2021-04-03 22:19:52  & 2070 & 1.38 & 1.0\\
& & & & & & & 2021-04-11 21:32:21 & 2390 & 1.32 & 1.2\\
& & & & & & & 2022-02-05 20:23:37 & 2160 & 1.20 & 1.3\\
3FHL\,J1055.6$-$0125 & 4FGL\,J1055.5$-$0125 & NVSS\,J105534$-$012617$^{\dagger}$ & 10 55 34.3 & $-$01 26 16 & SALT/RSS & 2.0  & 2021-06-04 17:04:09 & 2250 & 1.20 & 1.2\\
3FHL\,J1233.7$-$0145 & 4FGL\,J1233.7$-$0144 & NVSS\,J123341$-$014426$^{\dagger}$ & 12 33 41.3 & $-$01 44 23 & VLT/FORS & 1.3 & 2022-04-07 04:51:04 & 4050 & 1.17 & 1.9\\
3FHL\,J1307.6$-$4259 & 4FGL\,J1307.6$-$4259 & 1RXS\,J130737.8$-$425940$^{\dagger}$ & 13 07 38.0 & $-$42 59 39  &  SALT/RSS & 2.0 &   2021-05-30 22:00:31  & 2250 & 1.33 &  0.9\\
& & & & & & & 2021-06-09 21:27:28 & 2250 & 1.32 & 1.2\\
3FHL\,J1323.0+2941 & 4FGL\,J1323.0+2941  & 4C\,+29.48$^{\dagger}$ & 13 23 02.4 & +29 41 34 & Keck/ESI &  1.0 &  2022-06-04 06:05:18  & 7200 & 1.01 &  0.7\\
3FHL\,J1440.6$-$3846 & 4FGL\,J1440.6$-$3846 & 1RXS\,J144037.4$-$384658$^{\dagger}$ & 14 40 37.8 & $-$38 46 55 & VLT/FORS & 1.3 & 2022-03-01 08:55:45 & 1200 & 1.03 & 0.7\\
3FHL\,J1544.9$-$6641 & 4FGL\,J1545.0$-$6642 & PMN\,J1544$-$6641 & 15 44 59.0 & $-$66 41 47 & VLT/FORS & 1.3 & 2022-03-10 09:05:11 & 1220 & 1.35 & 0.5\\
3FHL\,J1607.9$-$2039 & 4FGL\,J1608.0$-$2038 & NVSS\,J160756$-$203942 & 16 07 56.9 & $-$20 39 42 & VLT/FORS & 1.3 & 2022-03-30 08:36:47 & 2430 & 1.02 & 0.6\\
3FHL\,J1637.8$-$3448 & 4FGL\,J1637.8$-$3449 & NVSS\,J163750$-$344915 & 16 37 51.0 & $-$34 49 15 & SALT/RSS  & 2.0 &   2021-06-05 01:09:53 & 2380 & 1.32 &  1.1\\
& & & & & & & 2021-06-12 16:49:51  & 2380 & 1.31 & 1.3\\
3FHL\,J1640.1+0629 & 4FGL\,J1640.2+0629 &  NVSS\,J164011+062827 & 16 40 11.0 & +06 28 27  &  Keck/ESI & 1.0 &  2022-06-04 08:19:10 & 6300 & 1.09 & 0.7\\
3FHL\,J1714.0$-$2028 & 4FGL\,J1714.0$-$2029 & 1RXS\,J171405.2$-$202747 & 17 14 05.4 & $-$20 27 52 & VLT/FORS & 1.3 & 2022-03-29 08:44:50 & 2430 & 1.01 & 0.6\\ 
& & & & & & & 2022-04-12 06:54:40 & 2430 & 1.06 & 1.0\\
& & & & & & & 2022-05-02 06:26:10 & 2430 & 1.01 & 1.2\\
3FHL\,J1754.1+3212 & 4FGL\,J1754.2+3212 &  RX\,J1754.1+3212$^{\dagger}$ & 17 54 11.8 & +32 12 23.1 & Keck/ESI & 1.0 & 2022-06-04 10:14:01  & 7200 &  1.03 &  0.9\\
3FHL\,J1823.6$-$3454 & 4FGL\,J1823.6$-$3453 & NVSS\,J182338$-$345412$^{\dagger}$ & 18 23 38.6 & $-$34 54 12 & VLT/FORS & 1.3 & 2022-03-11 09:06:16 & 500 & 1.24 & 0.6\\
  & & & & & & & 2023-06-20 05:35:36 & 900 & 1.02 & 1.3\\
3FHL\,J1849.2$-$1647 & 4FGL\,J1849.2$-$1647 & NVSS\,J184919$-$164723$^{\dagger}$ & 18 49 19.6 & $-$16 47 35 & VLT/FORS & 1.3 & 2022-04-12 07:51:09 & 1620 & 1.14 & 1.1\\
& & & & & & & 2022-05-01 08:52:14 & 2430 & 1.01 &  0.8 \\
3FHL\,J1925.0+2815 & 4FGL\,J1925.0+2815 &  NVSS\,J192502+28154 & 19 25 02.2 & +28 15 42  &  Keck/ESI & 1.0 &   2021-10-15 05:06:22 &  6300 & 1.09 & 1.1\\
3FHL\,J1942.7+1033 & 4FGL\,J1942.7+1033 & 1RXS\,J194246.3+103339$^{\dagger}$  & 19 42 47.5 & +10 33 27  &  Keck/ESI & 1.0  &  2022-06-04 12:27:43 & 7200 & 1.02 &  0.9\\
3FHL\,J1944.4$-$4523 & 4FGL\,J1944.4$-$4523 & 1RXS\,J194422.6$-$452326 & 19 44 22.4 & $-$45 23 33 & VLT/FORS & 1.3 & 2022-03-31 09:35:50 & 500 & 1.20 & 0.6\\
3FHL\,J1944.9-2143 & 4FGL\,J1944.9$-$2143 & NVSS\,J194455$-$214320 & 19 44 55.2 & $-$21 43 19  &  SALT/RSS & 2.0 &   2021-06-05 22:31:14 &  2250 & 1.21 &  1.2\\
& & & & & & & 2021-07-10 20:07:14 &  2250 & 1.23 &  1.3\\
& & & & & & & 2021-07-12 01:18:38 &  2250 & 1.26 &  1.3\\
3FHL\,J2247.9+4413 & 4FGL\,J2247.8+4413 & NVSS\,J224753+441317$^{\dagger}$ & 22 47 53.2 & +44 13 15  &  Keck/ESI &  1.0 &  2021-10-15 07:08:44 &  6300 & 1.11 &  0.8\\
3FHL\,J2304.7+3705 & 4FGL\,J2304.6+3704 & 1RXS\,J230437.1+370506$^{\dagger}$ & 23 04 36.7  & +37 05 07  &  Keck/ESI &  1.0 &  2021-10-15 09:06:21 &  6300 & 1.16 &  0.8\\

\hline
\hline

\end{tabular}
\tablefoot{The columns contain:  (1) 3FHL Name, (2) 4FGL Name, (3) Source Name, a $^{\dagger}$ indicates the source is in BZCat, (4) Right Ascension (J2000), (5) Declination (J2000), (6) Telescope and Instrument, (7) Slit Width in arcsec, (8) Start Time of the observations, (9) Exposure Time, (10) Average Airmass, and (11) Average Seeing.}
\end{sidewaystable*}

\begin{table*}
\caption{\label{tab_spgraphtech} Technical specifications of the spectrographs used in this study.
}

\centering
\begin{tabular}{lcccc}
\hline\hline

Instrument Name & Spectroscopic mode &  Wavelength coverage (\AA) & Throughput $p$ & Spectral resolution $\lambda$ / $\Delta \lambda$ \\

\hline

Keck/ESI  & Echellette & 3900 $-$ 10000 & $p \geq$ 28\% & $\sim$ 10000 \\
SALT/RSS & Long slit & 4500 $-$ 7500  & $p$ > 20\% & $\sim$ 1000 \\
VLT/FORS & Low resolution, GRISM 600RI+19 & 5000 $-$ 8500  & 20\% < $p$ < 30\% & $\sim$ 800 \\
VLT/FORS & Low resolution, GRISM 300I+11 & 6000 $-$ 11000  & 15\% < $p$ < 20\% & $\sim$ 400 \\

\hline
\end{tabular}
\end{table*}

\section{Redshift measurement and estimation of the total emission of blazars}\label{zmeas}
To measure the redshifts of the blazars, we searched for the emission or absorption lines in their optical spectra. For a reliable redshift determination, we required at least two independent features with consistent redshift values. The lines we looked for are listed in Section 2. Once we identified these features in the spectra, we calculated the EW by fitting the continuum with cubic splines and integrating the flux over each pixel. The uncertainties in the EW were estimated by summing, in quadrature, the errors from the normalized flux and the continuum placement \citep[see][]{Sem92}. Tables \ref{tabeqw}, \ref{tabeqw2} and \ref{tabeqw3} present the measured EWs along with their corresponding errors. 

To estimate the uncertainty in the redshift measurement, we considered the quadratic sum of the uncertainties from the wavelength calibration and the position of the detected features, the latter determined by a Gaussian fit. The relative error in wavelength calibration is estimated to be 6$-$12 $\times$ $10^{-5}$ (see details in Paper I and Paper II). The total estimated relative errors from the two contributions are between 1$-$5 $\times$ 10$^{-4}$. The redshift measurements and their error estimation are given in Table \ref{tabres1}.

The optical spectra of BL Lac objects result from both non-thermal emission, driven by jet activity, and thermal emission, originating from the stellar light of the elliptical host galaxy. Consequently, we modeled the spectral energy distribution (SED) by combining a power law ($f_{\lambda} \propto \lambda^{\alpha}$), representing the jet emission, with an elliptical galaxy template, representing the host galaxy \citep{Man01, Bru03}. Each spectrum was modelled with a single template. The fit required only two free parameters: the power-law slope and the jet-to-galaxy ratio. When an emission feature was visible in the spectrum, it was masked during the fit and added afterwards as a Gaussian component \citep[see, e.g.][]{Pit14}. We provide the results of the fits in Table \ref{tabres1}.

We also calculated the absolute magnitude of the detected host galaxies. To estimate slit losses, we assumed the effective radius of the host galaxy, $r_{e}$, to be 10 kpc, based on a de Vaucouleurs profile \citep{1948AnAp...11..247D, 1953MNRAS.113..134D}. Using the template spectra, we computed K-corrections without applying any evolutionary corrections. For hosts that were not detected, we fit the spectra with a power law, normalized at the center of the band. Due to the high S/N values of the spectra, very small relative errors are observed in the fitted parameters, approximately 10$^{-3}$. However, residual curvature seen in the spectra suggests either intrinsic changes in slope or possible calibration issues (such as flux calibration or extinction corrections). To account for this, we separately fit the red and blue halves of the spectra and used the difference between the blue and red parameters as the 1-sigma errors for the overall spectrum parameters. The uncertainties listed for the slope in Table \ref{tabres1} represent three times these values, or 3-sigma errors.

\begin{table*}
\caption{\label{tabeqw} Equivalent Widths in \AA~of the main absorption features detected in the spectra at the measured redshift for each source. }
\centering
\begin{tabular}{lccccc}
\hline\hline

Source Name &  CaHK & CaIG & Mgb & CaFe & NaID  \\
            &  $\lambda$ 3933.7 \& 3968.5 &  $\lambda$ 4304.4  & $\lambda$ 5174 & $\lambda$ 5269     & $\lambda$ 5892.5    \\
 (1)  &  (2)   &  (3) & (4) & (5) &  (6)  \\        

\hline
RX J0819.2$-$0756 & 3.7 $\pm$ 0.5  & 1.5 $\pm$ 0.2  & 2.6 $\pm$ 0.2 & 1.0 $\pm$ 0.1 &  1.3 $\pm$ 0.1  \\
1RXS J144037.4-384658 &  2.5$\pm$ 0.4 &  ---     &  1.6$\pm$ 0.2  &   ---  &    1.3$\pm$ 0.2 \\
PMN J1544-6641 & ---    &   0.6$\pm$0.2  &   1.1$\pm$0.2 & 0.9$\pm$0.2 &  1.5$\pm$0.2 \\
1RXS J171405.2-202747 & 3.2$\pm$0.6 & 0.7$\pm$0.2  &  1.3$\pm$0.2 & 0.8$\pm$0.2   & --- \\
NVSS J182338-345412-Gr600 &   ---    &      ---   &   0.4$\pm$0.1   & ---   &    0.3$\pm$0.1 \\
NVSS J182338-345412-Gr300 &  ---   &    ---    &   ---      &   ---     &  0.6$\pm$0.1 \\
NVSS J184919-164723        &  ---   &    ---    &  3.0$\pm$0.6   & ---    &   3.8$\pm$0.7 \\
NVSS J192502+28154     &     ---    &    ---    &  1.8$\pm$0.2  & 0.9$\pm$0.2  & 2.0$\pm$0.3 \\
1RXS J194422.6-452326    &   ---     &    ---    &  1.0$\pm$0.2  & 0.6$\pm$0.1  & 0.9$\pm$0.1 \\
NVSS J194455-214320   &    3.7$\pm$0.4  & 1.5$\pm$0.2  &   ---   &       ---     &   --- \\
\hline
\hline
\end{tabular}
\tablefoot{The NaID feature of 1RXS J194422.6-452326 is likely contaminated by telluric H$_{2}$O. The columns are (1) Source Name, (2) Equivalent Width of the CaHK feature with errors, (3) Equivalent Width of the CaIG feature with errors, (4) Equivalent Width of the Mgb feature with errors, (5) Equivalent Width of the CaFe feature with errors, (6) Equivalent Width of the NaID feature with errors. If the feature is not detected, the legend is `---'.}
\end{table*}

\begin{table*}
\small
\caption{\label{tabeqw2} Equivalent width in \AA~of the main emission features detected in the spectra at the measured redshift. For NVSS J182338-345412, two separate FORS observations were performed, one using Grism 600RI and the other using Grism 300I. The emission features detected in the spectrum of 1RXS J194422.6-452326  are in Table \ref{1RXSJ1944tab}.}

\centering
\begin{tabular}{lcccccc}
\hline\hline
Source Name & MgII &     [OII]     &      [OIII]                 &     [OIII]    & H$\alpha$ & [NII]$_{\rm b}$  \\
    &  $\lambda$ 2800   &  $\lambda$ 3727 &    $\lambda$ 4959 &   $\lambda$ 5007  & $\lambda$ 6562.8  & $\lambda$ 6583.6                    \\
    (1)  &  (2)   &  (3) & (4) & (5) &  (6)  & (7)\\  
\hline
PMN J0622-2605  & ---  & 2.9$\pm$0.2  & 1.6$\pm$0.2  &  4.3$\pm$0.2  &   ---  &      ---   \\         
4C+29.48    &  11.7$\pm$1.5   &     3.3$\pm$0.2  &   ---     &     ---    &     --- &       ---   \\
NVSS J160756-203942 & --- & 0.8$\pm$0.2   &  ---     &   0.8$\pm$0.2   &  ---   &     ---   \\
NVSS J182338-345412-Gr600 & --- &  ---   &    ---    &  0.3$\pm$0.1 &  ---  & 0.4$\pm$0.1 \\
NVSS J182338-345412-Gr300 & --- &  ---   &    ---    &  --- &  0.5$\pm$0.1 & 1.0$\pm$0.1 \\

\hline
\hline
\end{tabular}
\tablefoot{The columns are (1) Source Name, (2) Equivalent Width of the MgII feature with errors, (3) Equivalent Width of the [OII]$\lambda$ 3727 feature with errors, (4) Equivalent Width of the [OIII]$\lambda$ 4959 feature with errors, (5) Equivalent Width of the [OIII]$\lambda$ 5007 with errors, (6) Equivalent Width of the H$\alpha$ feature with errors, (7) Equivalent Width of the [NII]$\lambda$ 6583 feature with errors. If the feature is not detected, the legend is  `---'.}
\end{table*}

\section{Sources and results}\label{sources}
For the 29 BL Lac sources listed in Table \ref{tabobs1}, spectroscopic redshifts were determined for 12 of them, ranging from 0.1636 to 1.1427. Furthermore, two lower limits on redshift ($z \geq 1.0196$ and $z \geq 1.4454$) were identified. In the following, we present a detailed discussion of the results for each source based on our observations.

\subsection{1RXS J032342.6\texorpdfstring{$-$}{-}011131}
\begin{figure*}
   \centering
 \includegraphics[width=6.8truecm,height=5.75truecm]{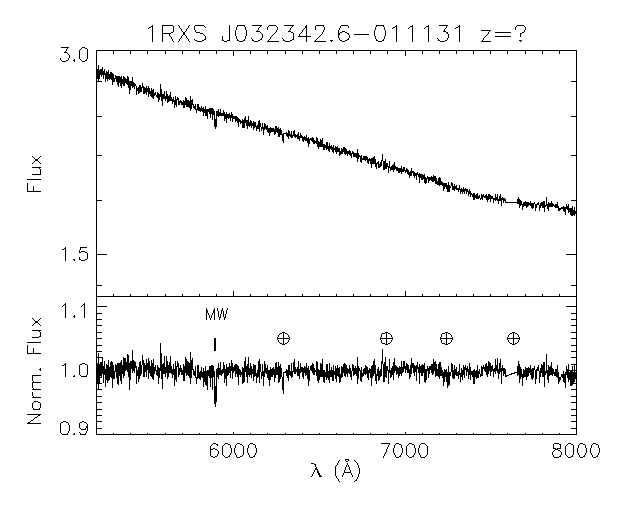}  \includegraphics[width=6.8truecm,height=5.75truecm]{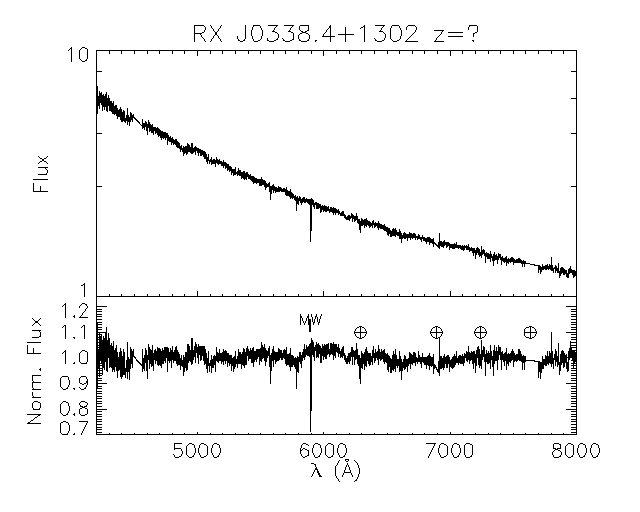} 
 \includegraphics[width=6.8truecm,height=5.75truecm]{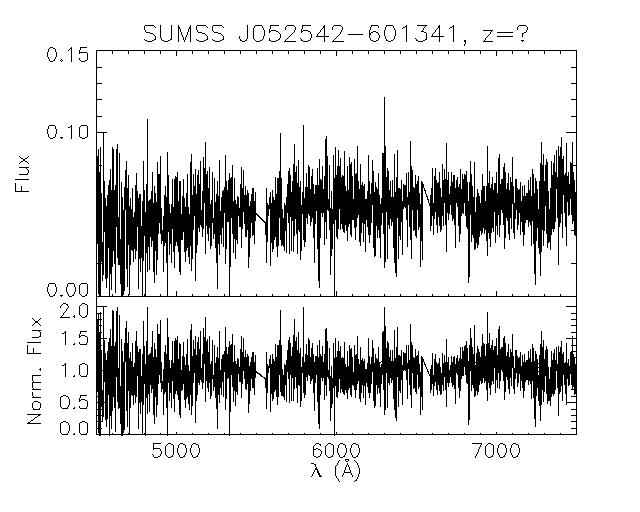}  \includegraphics[width=6.8truecm,height=5.75truecm]{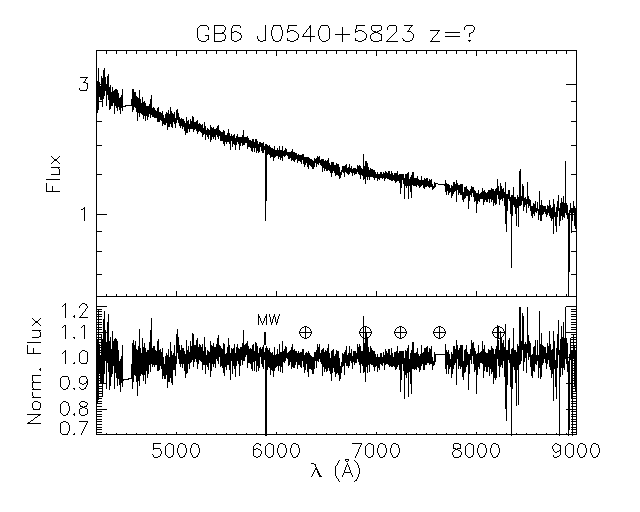}
  \includegraphics[width=6.8truecm,height=5.75truecm]{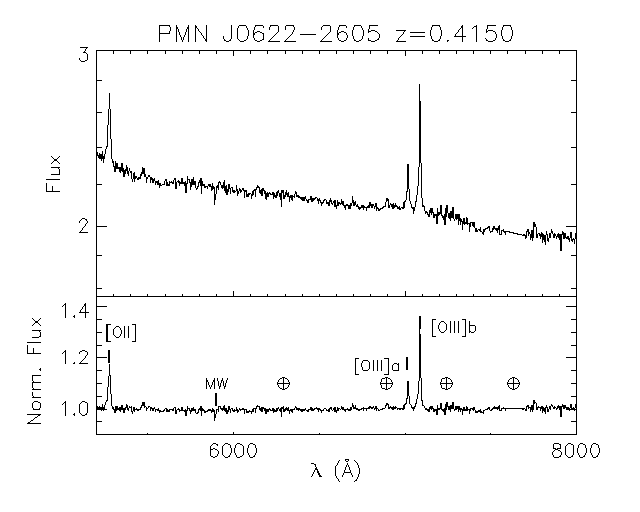}  \includegraphics[width=6.8truecm,height=5.75truecm]{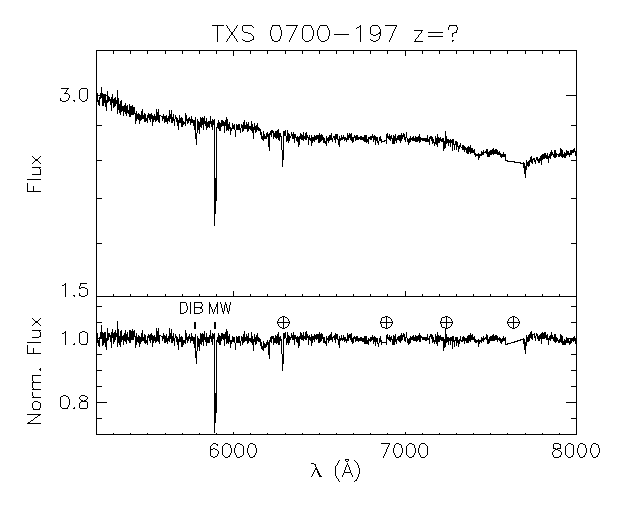}
 \includegraphics[width=6.8truecm,height=5.75truecm]{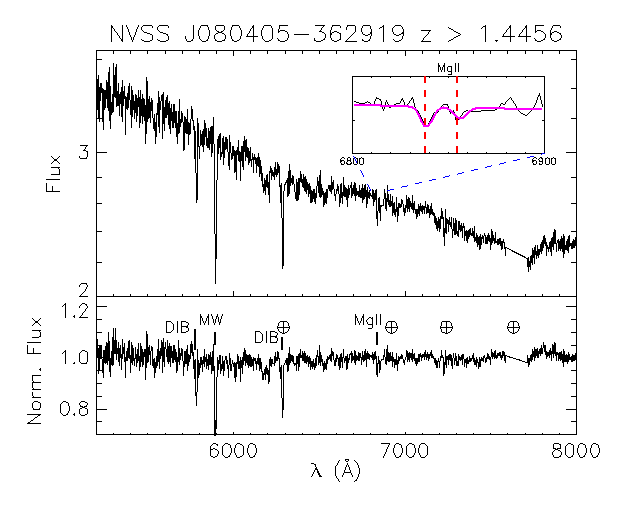}  \includegraphics[width=6.8truecm,height=5.75truecm]{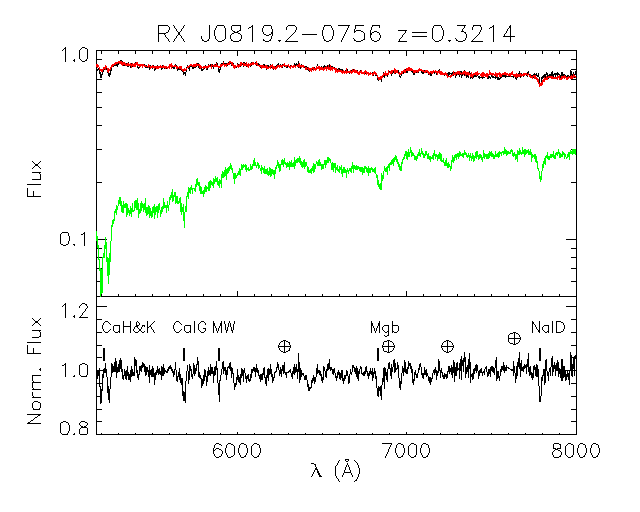}
   \caption{
   Flux-calibrated and normalized spectra of the first eight sources in Table \ref{tabobs1}. Each panel contains the spectrum and in some cases also a continuum and galaxy model for each source.  Each panel has  two parts. Upper: Flux-calibrated and telluric-corrected spectrum (black) alongside the best fit model (red). The flux is in units of 10$^{-16}$ erg  cm$^{-2}$ s$^{-1} $\AA$^{-1}$. The elliptical galaxy component is shown in green. We note that the plots are in observer frame and in logarithmic units, therefore different parts of the galaxy template are visible in each plot.
   The fits to intervening absorption systems are shown in magenta. Lower:  Normalised spectrum with labels for the detected spectral features. Atmospheric telluric absorption features are indicated by the symbol $\oplus$ and Galactic absorption features are labelled `MW' and 'DIB'. }

\label{fig_spec1}
    \end{figure*}

The SDSS spectrum of 1RXS J032342.6$-$011131 have been previously presented in \cite{2013ApJ...764..135S} and \cite{Pena20}, but no definitive redshift determination was reported in these studies. We observed this source using VLT/FORS on 23 September 2022 with a total exposure time of 1600 s, obtaining a high S/N ($\sim$132). The resulting spectrum is featureless and the redshift remains undetermined, as shown in the top left panel of Figure \ref{fig_spec1}.

\subsection{RX J0338.4\texorpdfstring{$+$}{+}1302}
  In Paper~II, a  Lick/KAST (KAST Double Spectrograph on the Shane 3-meter telescope at Lick observatory) observation was performed, with an exposure time of 7200 s, achieving a S/N of 32. The identification of the MgII doublet set a lower redshift limit for this source at $z \geq 0.3821$. This feature was also noted in earlier spectra reported by \cite{2016A&A...596A..10M, Pai17a}. We subsequently observed the source on 15 October 2021 with Keck/ESI using a 7200 s exposure time, achieving a moderate S/N of 60. The Keck/ESI observation revealed a featureless spectrum without additional identifiable features, as shown in the top right panel of Figure \ref{fig_spec1}. Note that the MgII absorber is not contained in the wavelength range of this spectrum.

\subsection{SUMSS J052542\texorpdfstring{$-$}{-}601341}
SUMSS J052542$-$601341 is a faint optical source with a magnitude of G = 19.90 $\pm$ 0.02 (Gaia-DR3; \citealt{2023A&A...674A...1G}). No spectral data for this source have been reported in the literature. Using photometric observations of the Lyman-alpha break from the {\em Swift} UV-Optical Telescope ({\em Swift}-UVOT; \citealt{2004ApJ...611.1005G}) and the Gamma-Ray Optical/Near-Infrared Detector (GROND; \citealt{2008PASP..120..405G}) on the MPG 2.2m telescope at ESO La Silla, \cite{2017ApJ...834...41K} estimated its photometric redshift to be 1.78. We obtained spectroscopic data for this source using SALT/RSS on 7 and 8 October 2021, with total exposure times of 2400 s and 1990 s, respectively. The resulting averaged spectrum, which is featureless, has a very low signal-to-noise ratio of 3 and is shown in the left panel of the second row in Figure \ref{fig_spec1}.  We note that the magnitude estimation we listed in Table \ref{tabres1} is equivalent to G = 21.5 $\pm$ 0.2, much weaker than the average Gaia magnitude. The reliability of the flux calibration in this particular case is however low due to the low signal-to-noise ratio of the spectrum.

\subsection{GB6 J0540\texorpdfstring{$+$}{+}5823}
GB6 J0540$+$5823 is identified as a hard $\gamma$-ray source and is located near the position of an IceCube\footnote{https://icecube.wisc.edu/} events \citep{2016MNRAS.457.3582P}. Previous spectra of this source were reported in \cite{2013ApJ...764..135S} using the Double Spectrograph on the 200-inch Hale Telescope at Mt. Palomar. A Spectrum of S/N of 25 was later observed with the 4-meter telescope at Kitt Peak National Observatory (KPNO) and presented in \cite{2019Ap&SS.364....5M}. \cite{2020MNRAS.497...94P} provided a S/N = 60 spectrum using the Gran Telescopio Canarias (GTC). In Paper~III, we presented spectra of this source obtained with the Lick/KAST instrument over a total exposure time of 8900 s and achieving a S/N of 34. None of the measured spectra revealed significant emission/absorption lines. Here, we present new observations conducted on 15 October 2021 with Keck/ESI, featuring a total integration time of 5900 s and S/N of 42, resulting in a featureless spectrum (see the right panel of the second row in Figure \ref{fig_spec1}).

\subsection{PMN J0622\texorpdfstring{$-$}{-}2605}\label{Dis_apen2}
The six-degree Field Galaxy Survey (6dF) spectrum of PMN J0622$-$2605, as reported in \cite{2009MNRAS.399..683J}, suggested a redshift of $z = 0.415$. However, the publicly available 6dF spectrum has a very low S/N. To improve the data quality, we observed this source using VLT/FORS on 6 February 2022, for a total exposure time of 1730 s and achieving a high S/N $\sim$ 90. Our observations detected the [OII], [OIII] $\lambda$ 4959, and [OIII] $\lambda$ 5007 emission lines, allowing us to firmly determine a precise redshift of $z = 0.4150 \pm 0.0001$. The H$\beta$ line could not be detected and we cannot set a significant upper limit on its flux as the line coincides with a strong atmospheric absorption feature.
This result confirms the 6dF redshift measurement despite the limitations of their low S/N spectrum. The spectrum of PMN J0622$-$2605 is shown in the left panel of the third row in Figure \ref{fig_spec1}. The [OIII] $\lambda$ 5007 line (see Figure \ref{fig_app_lines}, center) also displays a broad blueshifted component usually attributed to an outflow \citep[see, e.g.][]{Sin22}. The line can be fitted with two Gaussian functions offset by -$540 \pm 50 $ km/s and with a full width at half maximum (FWHM) $360$ km/s (core component at the systemic redshift) and $450$ km/s (blueshifted/outflow component). These values are quite typical for [OIII] lines in optical AGNs, both Type 1 and Type 2 \citep{Mul13} and do not allow to refine the classification of this source.

\begin{table}
\caption{\label{tabeqw3} Equivalent width in \AA~of the emission lines detected in 1RXS J194422.6-452326.}
\begin{center}
\begin{tabular}{cc}
\hline
\hline
  Line &  Equivalent Width \\
          &     \AA     \\    
\hline
 {[OIII]$\lambda$ 4959}    &   0.5 $\pm$ 0.2   \\
 {[OIII]$\lambda$ 5007}    &   1.1 $\pm$ 0.2   \\
 {[OI]$\lambda$ 6300}      &   1.1 $\pm$ 0.3   \\
 H$\alpha$                 &   2.3 $\pm$ 0.1   \\
 {[NII]$\lambda$ 6548}     &   1.0 $\pm$ 0.1   \\
 {[NII]$\lambda$ 6583}     &   3.1 $\pm$ 0.1   \\
\hline
\hline

\end{tabular}
\end{center}
\label{1RXSJ1944tab}
\end{table}%

\begin{table*}
\caption{\label{tabres1} Analysis results for the observed sources. Two separate FORS observations of NVSS J182338-345412 were performed, one using Grism 600RI and the other using Grism 300I. In these spectra absorption features of the host galaxy have been detected but their S/N was too weak to allow formal fitting with a galaxy model. We therefore fit the spectra with a power law and roughly estimated the host galaxy magnitude from the intensity of the features. The spectral bin width is 1\,\AA~for sources observed with Keck/ESI and SALT/RSS and 1.66\,\AA~for sources observed with VLT/FORS.}
\centering
\begin{tabular}{lccccccc}
\hline\hline

 Source Name  & S/N &  R$_{\rm c}$(BL Lac) & Redshift   & Flux Ratio  & R$_{\rm c}$(gal) & M$_{\rm R}$ & Slope   \\
              &     &   (obs)              &            &            &  (fit)            &   (gal)     &         \\  
   (1)  & (2) & (3)    &  (4)  &  (5)   &  (6)      &  (7)   &  (8) \\      
\hline 
1RXS J032342.6-011131  & 132 & 17.4 $\pm$ 0.1   &   ---      &          ---    &      ---    &   ---  & -1.2$\pm$ 0.1  \\  
RX J0338.4+1302    &     60  & 17.6 $\pm$ 0.2    &  ---      &       ---     &     ---   &    ---  & -2.5$\pm$0.4   \\  
SUMSS J052542–601341   &  3 & 21.2 $\pm$ 0.1   &   ---        &        ---     &     ---    &   ---  & -0.8$\pm$1.0 \\  
GB6 J0540+5823  & 42  & 16.8 $\pm$ 0.3   &  ---    &   ---      &   ---    &   ---  & -1.3 $\pm$ 0.1 \\   
PMN J0622-2605    & 89  & 17.5 $\pm$ 0.1  & 0.4150 $\pm$ 0.0001    &   ---    &   ---    &   ---   & -0.4 $\pm$ 0.1  \\  
TXS 0700-197      &    110  & 16.7$\pm$ 0.1  &    ---     &      ---   &   --- &      ---  & -0.4 $\pm$ 0.1  \\  
NVSS J080405-362919   &  70 & 17.1 $\pm$ 0.1 & $>$1.4454 $\pm$ 0.0003  &     ---     &     ---    &   ---  & -1.0 $\pm$ 0.1 \\  
RX J0819.2-0756   &  61 & 18.5 $\pm$ 0.2 &  0.3214 $\pm$ 0.0002  &   1.8 $\pm$ 0.6  &  19.4 $\pm$ 0.4  & -22.0  &-1.2 $\pm$ 0.2 \\  
NVSS J084121-355506  &    96  &16.7 $\pm$ 0.2  & $>$1.0196 $\pm$ 0.0001 &    ---    &    ---   &   ---  & -1.2 $\pm$ 0.1   \\
1RXS J094709.2-254056  & 126  & 16.2 $\pm$ 0.1   &  ---      &     ---    &    ---   &   ---  & -0.9 $\pm$ 0.2   \\  
NVSS J105534-012617  &  79 & 18.2 $\pm$ 0.1  &    ---     &     ---      &  ---    &  ---   & -1.0 $\pm$ 0.2   \\
NVSS J123341-014426   & 144 & 18.0 $\pm$ 0.1    &  ---      &   ---     &  ---    &   --- &  -0.7 $\pm$ 0.1  \\  
1RXS J130737.8-425940 & 109  & 16.7 $\pm$ 0.1  &    ---    &     ---     &    ---    &   ---   & -1.1 $\pm$ 0.1   \\  
4C +29.48    &   43 & 19.7 $\pm$ 0.2  & 1.1427 $\pm$ 0.0001    &  ---     &     ---    &   ---  &  -0.4 $\pm$ 0.2  \\  
1RXS J144037.4-384658  & 75 & 17.7 $\pm$ 0.2 &  0.3583 $\pm$ 0.0002   &  4.8 $\pm$ 0.9   &  19.5 $\pm$ 0.2 & -22.3  & -1.2 $\pm$ 0.1 \\  
PMN J1544-6641  &  70   & 17.7 $\pm$ 0.1  & 0.2595  $\pm$ 0.0004   &  4.9  $\pm$ 0.3    & 19.1  $\pm$ 0.1 & -21.7  & -1.1  $\pm$ 0.1  \\  
NVSS J160756-203942  &   62   & 18.3  $\pm$ 0.1  &  0.5479  $\pm$ 0.0001   &    ---    &    ---   &    ---   & -0.5  $\pm$ 0.1 \\  
NVSS J163750-344915  &   93  & 16.8 $\pm$ 0.1  &  ---   &     ---    &    ---    &   ---  & -0.9 $\pm$ 0.1 \\ 
NVSS J164011+062827   & 112  & 17.7 $\pm$ 0.2  &    ---    &      ---     &     ---     &  ---  &  -0.9 $\pm$ 0.1 \\ 
1RXS J171405.2-202747 &   65  & 17.6 $\pm$ 0.1  & 0.5222 $\pm$ 0.0002   &  3.5 $\pm$ 0.5   &  19.4 $\pm$ 0.2  & -23.8   & -1.4 $\pm$ 0.1\\   
RX J1754.1+3212    &    170  & 16.5 $\pm$ 0.2  &    ---      &       ---     &   ---    &   ---   & -1.1$\pm$0.04 \\  
NVSS J182338-345412-Gr600   & 110  & 15.8 $\pm$ 0.2  & 0.1826 $\pm$ 0.0002    &   ---      &   $\sim$17.8    & $\sim$-22.1 & -1.2$\pm$0.1 \\
NVSS J182338-345412-Gr300    & 98  &15.8 $\pm$ 0.2 &  0.1826 $\pm$ 0.0002   &    ---      &    ---    &    ---  & -1.4$\pm$0.3 \\ 
NVSS J184919-164723  &   46 & 19.0 $\pm$ 0.1 &  0.3485 $\pm$ 0.0005   &  2.0 $\pm$ 0.4  & 19.9 $\pm$ 0.2  & -21.8  & -1.0 $\pm$ 0.1 \\
NVSS J192502+28154    &  65 & 16.5 $\pm$ 0.1  & 0.1636 $\pm$ 0.0003     &  ---      &    17.6 $\pm$ 0.4   &-22.2 & -1.3$\pm$ 0.1  \\ 
1RXS J194246.3+103339  & 97 & 16.0 $\pm$ 0.2     & ---      &    ---     &    ---   &     ---  & -0.7$\pm$0.1 \\
1RXS J194422.6-452326  &  75  & 17.2 $\pm$ 0.1 &  0.2358 $\pm$ 0.0002  & 5.3 $\pm$ 0.6  & 18.2 $\pm$ 0.1 & -22.3 & -1.1 $\pm$ 0.1 \\
NVSS J194455-214320   &  76  & 17.8 $\pm$ 0.1 & 0.4263 $\pm$ 0.0003 & 1.9 $\pm$ 0.1  & 18.8 $\pm$ 0.1 &  -23.6 & -1.1 $\pm$ 0.2 \\ 
NVSS J224753+441317   & 133  & 16.6 $\pm$ 0.3  &    ---  &    ---   &     ---   & ---   & -1.4  $\pm$ 0.1 \\
1RXS J230437.1+370506   & 95 & 17.9 $\pm$ 0.3    &  ---   &     ---     &     ---     &  ---  & -1.5 $\pm$ 0.1 \\
\hline
\hline
\end{tabular}
\tablefoot{The columns are (1) Source Name; (2) Median signal-to-noise ratio per spectral bin measured in continuum regions; (3) R$_{\rm c}$, Cousins magnitude of the BL Lac spectrum corrected for Galactic reddening, telluric absorption, and slit losses with errors. Slit losses were estimated using an effective radius r$_e$ = 10 kpc for all sources; (4) Redshift or lower limit with errors; (5) Flux Ratio jet/galaxy at 5500 \AA~in rest frame; (6) R$_{\rm c}$, Cousins magnitude of the galaxy with the same corrections as in column (3); (7) Absolute R Magnitude of the galaxy, the errors are the same as those in column (6); (8) Power-Law Slope $\alpha$ with f$_{\lambda} \propto \lambda^{\alpha}$ with errors. For $\alpha$, 3-sigma errors are shown; all other quantities are shown with 1-sigma errors. If the entry is unknown, the legend is `---'.

}
\end{table*}

\subsection{TXS 0700\texorpdfstring{$-$}{-}197}
A medium S/N spectrum of TXS 0700$-$197 was analyzed using the Low Resolution Imaging Spectrograph at the W. M. Keck Observatory, as reported by \cite{2013ApJ...764..135S}. No significant spectral lines were detected in the observations. On 6 February 2022, we observed this source with VLT/FORS for a total exposure time of 2595 s and achieving a high S/N $\sim$ 110. The resulting spectrum, shown in the right panel of the third row in Figure \ref{fig_spec1}, is featureless except for Galactic features such as NaID and the $\lambda$5780 diffuse interstellar band (DIB).

\subsection{NVSS J080405\texorpdfstring{$-$}{-}362919}
NVSS J080405$-$362919 is a highly absorbed source (E$_{B-V}$ = 1.04) located near the Galactic plane (b $= -2.8$). No prior spectral data for this source are available in the literature. Since it is faint (G = 19.7 $\pm$ 0.01) and there is confusion with nearby sources, no photometric monitoring is available. We made observations using VLT/FORS on 10 March 2022 and 2 May 2022, with a total exposure time of 2430 s for each session and achieving a S/N of 70. The spectrum (left panel of the fourth row in Figure \ref{fig_spec1}) displays a strong Galactic NaID line (EW = $4.0 \pm 0.1$ \AA), furthermore the strong Galactic DIBs $\lambda$5780 (EW = $1.5 \pm 0.1$ \AA) and $\lambda$ 6284 (EW = $2.2 \pm 0.1$ \AA) \citep[see e.g.][]{Jen94} are also visible.
In addition, we clearly detect a weaker extragalactic feature: a MgII absorber with EW = $0.80 \pm 0.14$ \AA~(as shown in the figure inset), establishing a lower redshift limit of $z \geq 1.4454$ for the source. The EWs of the two components are $0.5 \pm 0.1$ \AA~and $0.3 \pm 0.1$ \AA. Their ratio is $1.7~\pm~0.4$, which indicates a mildly saturated system \citep{Spitzer78}.

\subsection{RX J0819.2\texorpdfstring{$-$}{-}0756}
RX J0819.2$-$0756 has been reported at redshift $z = 0.85115$ in the 6dF survey \citep{2009MNRAS.399..683J}. Upon examination, however, the 6dF spectrum appears featureless. Furthermore, a photometric redshift of $z = 0.37$ was suggested by \cite{3HSP}. \cite{2016AJ....151...95A} conducted a short observation of this source ($\sim$1200 s) using the IMACS medium-resolution spectrograph on the 6.5 m Magellan telescope at Cerro Manqui, Chile. However, the S/N = 13 spectrum was featureless. In Paper~II, we observed this source with the NTT/EFOSC for 6650 s, obtaining a moderate S/N ($\sim$ 36) spectrum. The tentative redshift was estimated at $z$ $\sim$ 0.320 based on the detection of CaHK lines. In this study, we observed RX J0819.2$-$0756 on 25 February 2022, using VLT/FORS with improved S/N of 61, over a total exposure time of 5190 s. Identification of absorption lines including CaHK, Ca I G, Mgb, and Na I D allowed us to firmly determine the redshift as $z = 0.3214 \pm 0.0002$. The spectrum of RX J0819.2$-$0756 is displayed in the right panel of the fourth row in Figure \ref{fig_spec1}.

\subsection{NVSS J084121\texorpdfstring{$-$}{-}355506}
\begin{figure*}
   \centering
 \includegraphics[width=6.8truecm,height=5.75truecm]{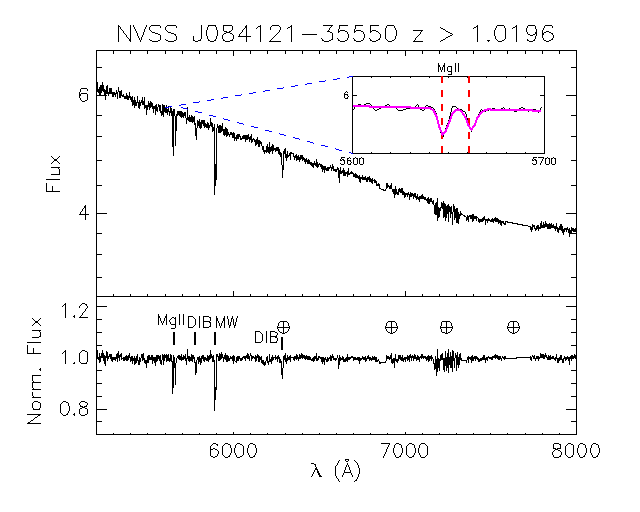}
\includegraphics[width=6.8truecm,height=5.75truecm]{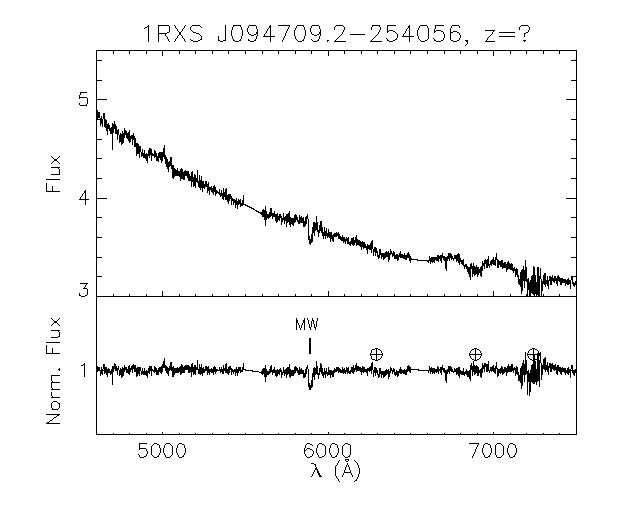}
 \includegraphics[width=6.8truecm,height=5.75truecm]{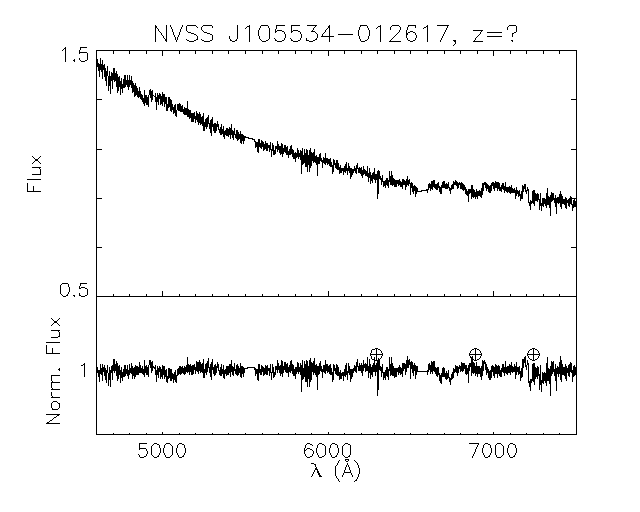}
 \includegraphics[width=6.8truecm,height=5.75truecm]{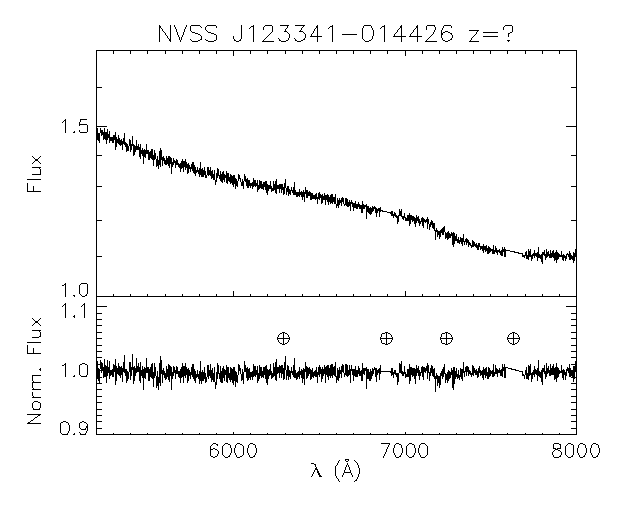}
 \includegraphics[width=6.8truecm,height=5.75truecm]{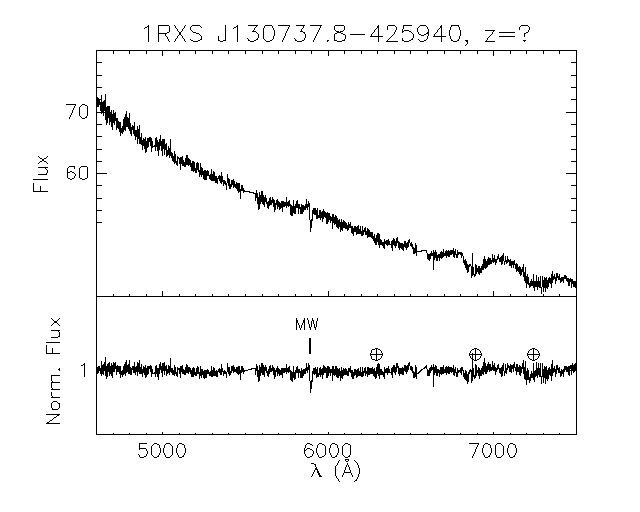}
 \includegraphics[width=6.8truecm,height=5.75truecm]{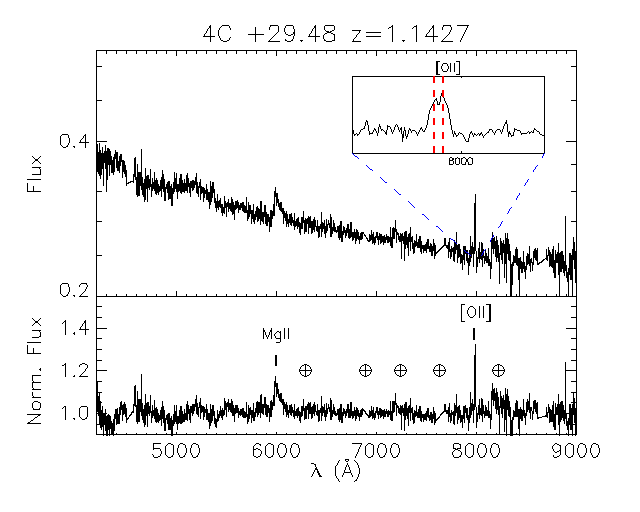}
 \includegraphics[width=6.8truecm,height=5.75truecm]{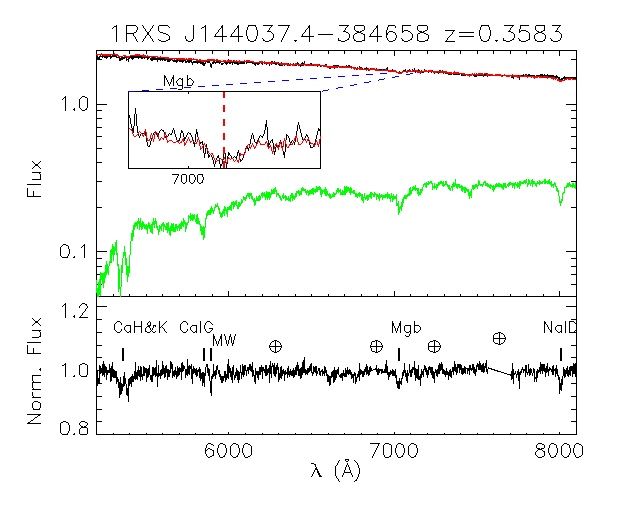}
 \includegraphics[width=6.8truecm,height=5.75truecm]{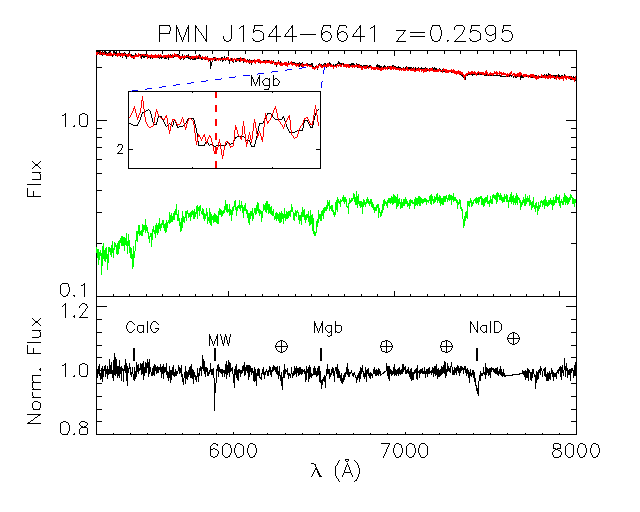}
 \caption{Same as Fig. \ref{fig_spec1} for sources 9 to 16 in Table \ref{tabobs1}.}
\label{fig_spec2}
    \end{figure*}

NVSS J084121$-$355506 is a heavily absorbed source (E$_{B-V}$ = 0.4249). A low S/N, featureless spectrum was previously obtained by \cite{2013A&A...559A..58M} using the New Technology Telescope (NTT) of the European Southern Observatory with an observing time of 3600 s, and medium S/N spectra were obtained with the NTT by \cite{2017AJ....153..157T}. On 4 March 2022, we observed this source using VLT/FORS, obtaining a S/N of 96 for a total integration time of 2595 s. Similar to the case of NVSS J080405$-$ 362919, strong $\lambda$ 5780 and $\lambda$ 6284 DIBs are present, but at lower intensity. Our observations reveal the presence of an MgII absorber with EW = $1.10 \pm 0.08$ \AA~ and establish a lower redshift limit of $z \geq 1.0196$. The EWs of the two components are $0.6 \pm 0.05$ \AA~and $0.5 \pm 0.06$ \AA~and their ratio is $1.2 \pm 0.1$, suggesting a fully saturated system \citep{Spitzer78}.
The source's spectrum is presented in the top left panel of Figure \ref{fig_spec2}.

\subsection{1RXS J094709.2\texorpdfstring{$-$}{-}254056}
The spectrum of 1RXS J094709.2$-$254056 was previously studied by \cite{2017Ap&SS.362..228P} using the Southern Astrophysical Research Telescope, with a 600 s exposure time, achieving a S/N of 95 and revealing no discernible features. We conducted further observations of this source using SALT/RSS on 3 April 2021, 11 April 2021, and 5 February 2022, with exposure times of 2070 s, 2390 s, and 2160 s, respectively, and achieved a high S/N $\sim$ 126. Neither our resulting spectrum show significant features. The spectrum is shown in the upper right panel of Figure \ref{fig_spec2}. 

\subsection{NVSS J105534\texorpdfstring{$-$}{-}012617}
A very low S/N, featureless spectrum of NVSS J105534$-$012617 was previously obtained using the 6dF survey by \cite{2009MNRAS.399..683J}. In a separate study, \cite{2020MNRAS.497...94P} obtained a high S/N GTC spectrum of the source, but it lacked any spectral characteristics. We observed this source with SALT/RSS on 4 June 2021, with a total exposure time of 2250 s and achieved a high S/N of 79. Our observations revealed no significant spectral features, as shown in the left panel of the second row in Figure \ref{fig_spec2}. The redshift of the source is still undetermined.

\subsection{NVSS J123341\texorpdfstring{$-$}{-}014426}
Spectra of NVSS J123341$-$014426 were previously examined by \cite{2013ApJ...764..135S} using data from the Marcario Low-Resolution Spectrograph on the Hobby-Eberly Telescope and by \cite{2014A&A...569A..95K} using data from SDSS. Both analyses reported featureless spectra and did not determine a spectroscopic redshift. Later, \cite{2020MNRAS.497...94P} observed the source with the GTC obtaining a S/N = 40 spectrum, but no significant spectral features were detected. In our study, we observed this source on 7 April 2022 using VLT/FORS with a 4050 s exposure, obtaining spectra of high S/N of 144 and found no prominent emission or absorption features. The corresponding spectrum is shown in the right panel of the second row in Figure \ref{fig_spec2}. The redshift of the source still remains undetermined.

\subsection{1RXS J130737.8\texorpdfstring{$-$}{-}425940}
A high S/N spectrum was obtained by \cite{2013A&A...559A..58M} using NTT/EFOSC2 with an exposure time of 1200 s, yielding a featureless spectrum. In our study, we observed the spectrum of 1RXS J130737.8$-$425940 using SALT/RSS on 30 May 2021, and 9 June 2021, with a total exposure time of 2250 s for each session and achieved a high S/N $\sim$ 109. The resulting spectrum, as shown in the left panel of the third row of Figure \ref{fig_spec2}, is featureless. The redshift of the source has not yet been determined.

\subsection{4C \texorpdfstring{$+$}{+}29.48}\label{Dis_apen1}

\citet{2018A&A...612A.109G} discussed the identification of the optical counterpart of 4C $+$29.48. The proposed object has been observed in SDSS DR12, and its redshift is reported as 1.142. However, due to the low S/N of the SDSS spectrum, we observed this source again using Keck/ESI on 4 June 2022, accumulating a total exposure time of 7200 s and obtaining a moderate S/N of 43. Our observations revealed the presence of an MgII emission line with an asymmetric profile extended towards long wavelengths and a weaker [OII] line at redshift $z \sim 1.143$. 
Profiles similar to the ones of the MgII line are often detected in the CIV lines of Radio Loud Quasars \citep{Wil95} and in the MgII lines of bright $\gamma$-ray blazars such as 3C 279 \cite[][see their Figure 1]{Puns20}. The asymmetry parameter A\footnote{A=$(\lambda_{25}-\lambda_{80})\over FWHM$ where $\lambda_{25}$ and $\lambda_{80}$ are the central wavelengths measured at 25\% and 80\% of the line flux level.} \citep{Wil95} of the MgII line of 4C $+$29.48 is 0.2 while it is 0.6 for 3C 279 indicating mild asymmetry. A possible explanation of this profile is a very strong jet combined with a weak accretion disk \citep{Puns20}. Careful examination of the [OII] line shows the presence of a double peak (Figure \ref{fig_app_lines}) which suggests that the doublet is resolved in our data. Indeed we could not fit the line with a single Gaussian ($\chi^2/dof =2.42$) but we could find a satisfactory fit with two Gaussians ($\chi^2/dof =1.67$). An F-test shows that this improvement in $\chi^2/dof$ has only 1.5\% probability of happening with three less degrees of freedom and we consider that the second component has been detected. The two Gaussians can be identified with the two components of the [OII] doublet: $\lambda$ 3726 and $\lambda$ 3729, at redshift $z = 1.1427 \pm 0.0001$, which we take as the systemic redshift. The intensity ratio is $\sim$1.3 which is consistent with a thin plasma at electron density n$_{\rm e} \sim 200 \,{\rm cm^{-3}}$ at T = 10$^{4}$ K, typical values of HII regions \citep[see Section 5.6 and Figure 5.8,][]{Oster06}. The full spectrum of 4C $+$29.48 is shown in the right panel of the third row of the Figure \ref{fig_spec2}.

\subsection{1RXS J144037.4\texorpdfstring{$-$}{-}384658}
No prior spectrum for 1RXS J144037.4$-$384658 has been reported in the literature. \cite{2020ApJ...898...18R} estimated its photometric redshift as $z_{phot} = 0.15^{+0.28}_{-0.06}$, using data from six {\em Swift}-UVOT filters as well as the SDSS $g^{'}$, $r^{'}$, $i^{'}$, and $z^{'}$ optical filters. These observations were conducted with the 0.65 m SARA-CTIO telescope in Chile and the 1.0 m SARA-ORM telescope in the Canary Islands. We observed this source on 1 March 2022, with VLT/FORS, obtaining a total exposure time of 1200 s and a high S/N of 75. Our analysis identified absorption lines such as CaHK, CaIG, Mgb and NaID (as shown in Figure~\ref{fig_spec2}, fourth row, left panel), which enabled us to determine a precise redshift of $z = 0.3583 \pm 0.0002$.

\subsection{PMN J1544\texorpdfstring{$-$}{-}6641}
The spectrum previously reported by \cite{Pena20}, observed with the Blanco 4-m telescope at Cerro Tololo Inter-American Observatory in Chile, was featureless. On 10 March 2022, we obtained the spectrum of PMN J1544$-$6641 using VLT/FORS, achieving a S/N of 70 with a total exposure time of 1220 s. Our analysis revealed absorption lines, including CaIG, Mgb, and NaID (shown in Figure \ref{fig_spec2}, fourth row, right panel), allowing us to determine a redshift of $z = 0.2595 \pm 0.0004$.

\subsection{NVSS J160756\texorpdfstring{$-$}{-}203942}
\begin{figure*}
   \centering
 \includegraphics[width=6.8truecm,height=5.75truecm]{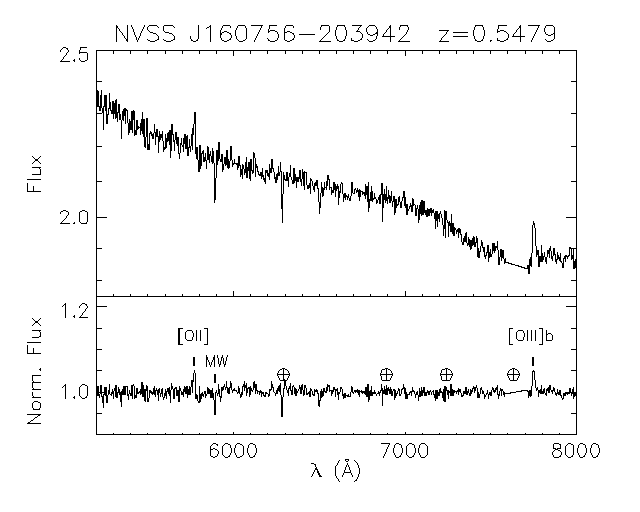}
 \includegraphics[width=6.8truecm,height=5.75truecm]{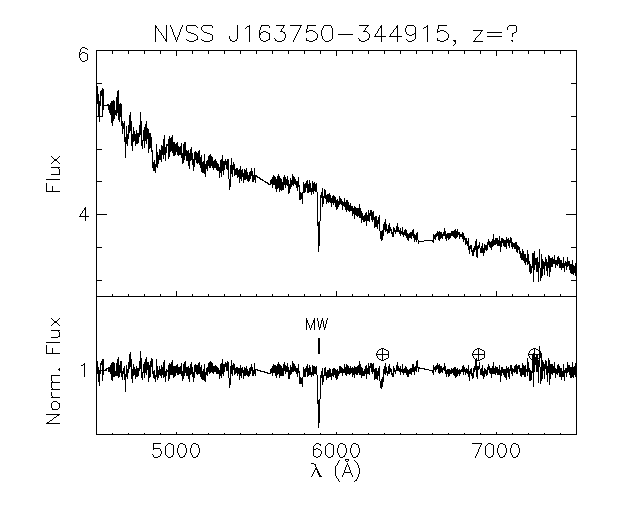}
 \includegraphics[width=6.8truecm,height=5.75truecm]{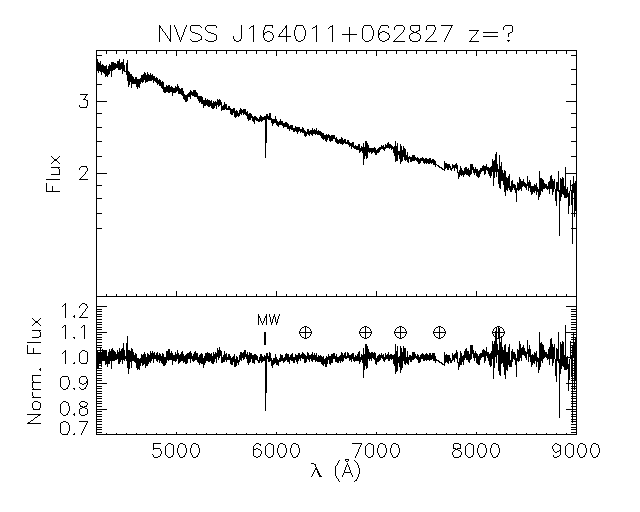}
 \includegraphics[width=6.8truecm,height=5.75truecm]{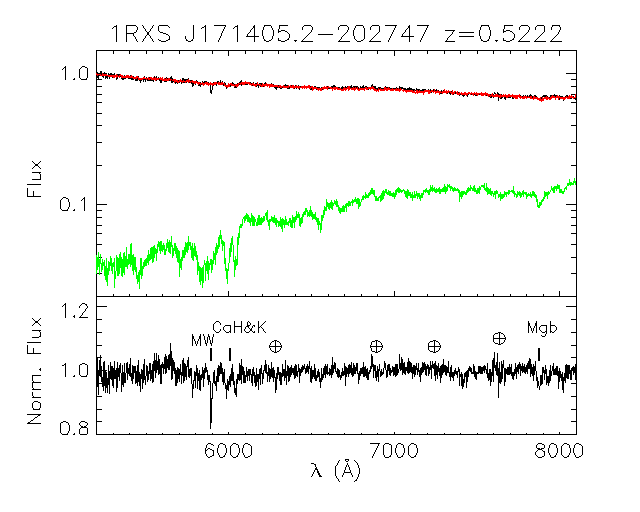}
 \includegraphics[width=6.8truecm,height=5.75truecm]{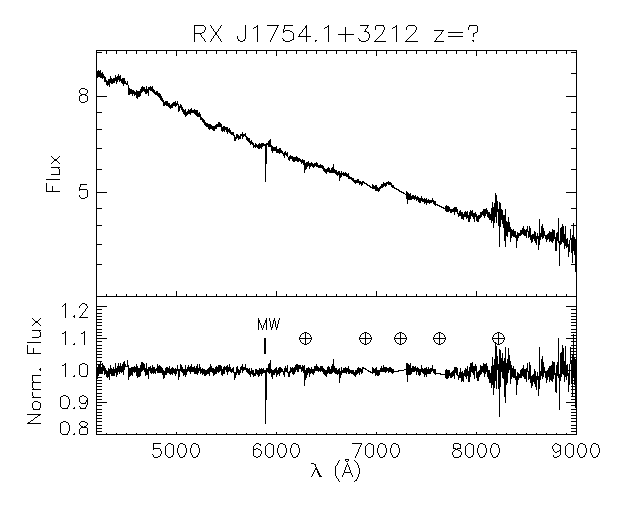}
 \includegraphics[width=6.8truecm,height=5.75truecm]{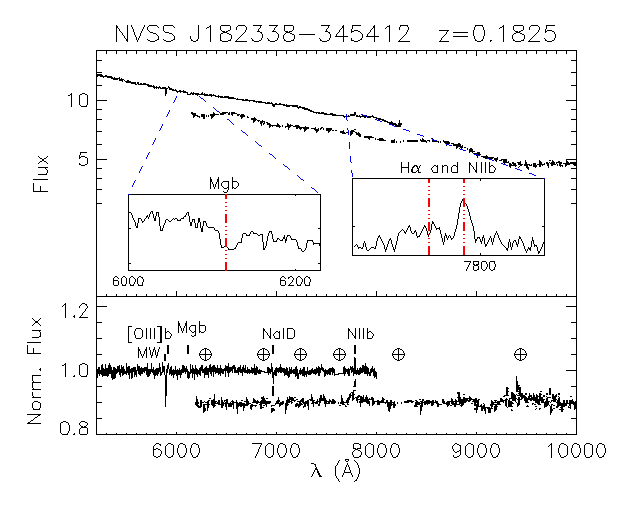}
 \includegraphics[width=6.8truecm,height=5.75truecm]{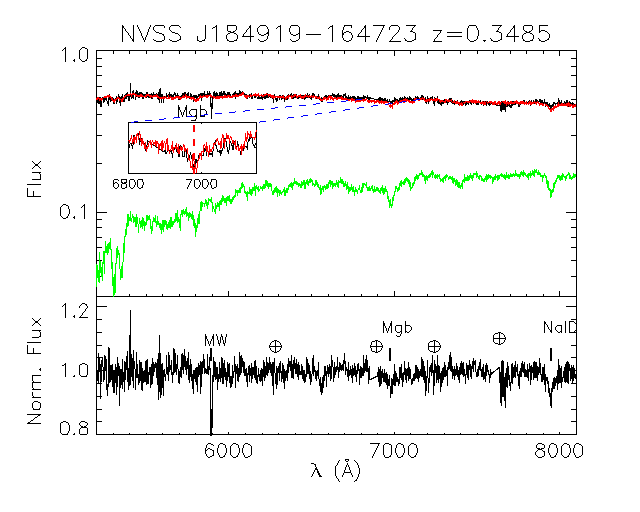}
  \includegraphics[width=6.8truecm,height=5.75truecm]{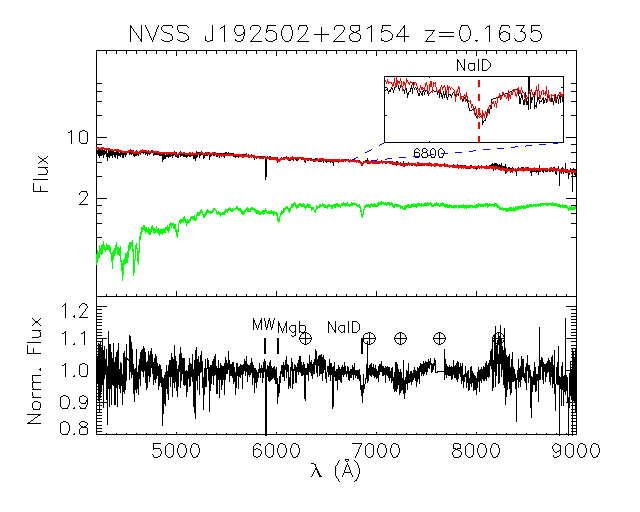}
 \caption{Same as Fig. \ref{fig_spec1} for sources 17 to 24 in Table \ref{tabobs1}.}
\label{fig_spec3}
    \end{figure*}

No previous spectrum has been known for NVSS J160756$-$203942. We conducted observations on 30 March 2022, using VLT/FORS with an exposure time of 2430 s, achieving a moderate S/N of 62. In our observed spectra, we detected [OII] and [OIII]b emission lines, enabling the determination of the redshift as $z = 0.5479 \pm 0.0001$. The resulting spectrum is presented in the top left panel of Figure \ref{fig_spec3}.

\subsection{NVSS J163750\texorpdfstring{$-$}{-}344915}
The spectrum of NVSS J163750$-$344915 was previously obtained by \cite{2017Ap&SS.362..228P} using the Goodman Spectrograph at Southern Astrophysical Research Telescope (SOAR), achieving a S/N of 60. The resulting spectrum was featureless. In Paper~I, we observed this source with EFOSC2, obtaining an improved S/N of 80, but no spectral lines were detected. In this study, we report the results of new observations conducted with SALT/RSS on 5 June and 12 June 2021, with a total exposure time of 2380 s per session, and obtained an even higher S/N of $\sim$ 93. The resulting spectra, shown in the top right panel of Figure \ref{fig_spec3}, remain featureless, and the redshift of the source is still undetermined.

\subsection{NVSS J164011\texorpdfstring{$+$}{+}062827}
No previous spectrum for NVSS J164011$+$062827 has been reported in the literature. On 4 June 2022, we observed this source using Keck/ESI with a total exposure time of 6300 s and achieved a very high S/N of 112, but no spectral features were detected (see Figure \ref{fig_spec3}, second row, left panel). 

\subsection{1RXS J171405.2\texorpdfstring{$-$}{-}202747}
No prior spectral studies of 1RXS J171405.2$-$202747 have been documented in the literature. A photometric redshift $z=0.09$ of its counterpart 3HSP J171405.4$-$202752 was reported in the 3HSP catalog \citep{3HSP} and on this basis \citet{Nievas22} classified the source as an Extreme High Energy BL Lac. However, the optical counterpart reported in the 3HSP catalog has a measurable parallax, $\pi$ = 0\farcs25 $\pm$ 0\farcs03 in the Gaia catalog \citep{2023A&A...674A...1G}.  This is not compatible with an extragalactic origin, and indeed the 4LAC catalog reports for 1RXS J171405.2$-$202747 another optical counterpart with negligible parallax. We report here on observations of the 4LAC counterpart.
We conducted observations of this source using VLT/FORS on three occasions: 29 March 2022, 12 April 2022, and 2 May 2022, with each observation having a total exposure time of 2430 s, obtaining a moderate S/N of 65. Our spectral analysis identified the presence of CaHK and Mgb absorption lines (as shown in Figure \ref{fig_spec3}, second row, right panel), and determined the redshift of the source to be $z = 0.5222 \pm 0.0002$.

\subsection{RX J1754.1\texorpdfstring{$+$}{+}3212}
\cite{2013ApJ...764..135S} presented spectra of RX J1754.1$+$3212 obtained with the LRIS spectrograph \citep{Oke95} at the W.~M.~Keck Observatory, but did not establish a definitive redshift. \cite{Shaw13b} triggered another spectral observation, unfortunately in poor observing conditions, when the source was in low state, yet the redshift remained undetermined. We examined the KAIT\footnote{\url{http://herculesii.astro.berkeley.edu/kait/agn/}} light curve of the source and determined that the source was at magnitude R $\sim$ 15.8 and R $\sim$ 17.0 during the first and second observations, respectively \citep{Li03}. 

We conducted observations of this source on 4 June 2022, using Keck/ESI with a total integrated exposure time of 7200 s, achieving a high S/N of 170. The source was again in low state at r $\sim$ 16.9 according to Zwicky Transient Facility (ZTF) monitoring \citep{Masc19}. Our spectra were featureless, and the redshift remains unknown. The resulting spectrum is shown in the left panel of the third row in Figure \ref{fig_spec3}.

\subsection{NVSS J182338\texorpdfstring{$-$}{-}345412}
A featureless spectrum of NVSS J182338$-$345412, was reported by \cite{2013A&A...559A..58M}, obtained using the NTT/EFOSC2 on September 5, 2012. We observed the same source on two occasions using VLT/FORS. First, on 11 March 2022 with a Grism 600RI and a GG435 filter (wavelength range $\sim$ 5120$-$8450 \AA), for a total exposure time of 500 s, achieving a high S/N of 110. In the corresponding spectrum, we identified a likely emission feature around 7805~\AA, which, when interpreted as [NII]b, suggests a redshift of $z \sim 0.183$. Other possible weaker features could be interpreted as Mgb, NaID, and [OIII]b at this redshift.
To better investigate this possibility, we performed a second observation on 20 June 2023 with a different configuration using Grism 300I and an OG590 filter (wavelength range $\sim$ 6000$-$11000 \AA) for a total exposure time of 900 s and achieving a high S/N $\sim$ 100. The flux calibrated spectra from both observations are presented in Figure \ref{fig_spec3} (third row, right panel) and a zoom between 6800 and 8100 \AA~of the normalized spectra is shown in Figure \ref{fig_app_lines}. We note that the different flux levels of the two spectra are likely due to long term variability, unfortunately, due to confusion issues, no reliable optical light curve of NVSS J182338$-$345412 is available to test this hypothesis.
The second spectrum confirms the results of the first, clearly displaying the $H_{\alpha}$-[NII] complex and the NaID absorption. Combining the two spectra we can establish a precise redshift for the source of $z = 0.1826 \pm 0.0002$.
Due to the weakness of the absorption features we were not able to fit the galaxy model to the spectra. We thus roughly estimated the galaxy magnitude from the intensity of Mgb and NaID to be M$_{\rm R}$ $\sim -22.1$ with a likely error of $\pm$ 0.5.

\subsection{NVSS J184919\texorpdfstring{$-$}{-}164723}
A low S/N spectrum of NVSS J184919$-$164723 was reported by \cite{2021ApJS..254...26R}, observed with the Blanco 4-m telescope, suggested a redshift z $>$ 0.64-0.67 for this source. We conducted observations of this source with VLT/FORS on 12 April 2022, and 1 May 2022, with total exposure times of 1620 s and 2430 s, respectively, achieving a moderate S/N of 46. Our spectra show the presence of Mgb and NaID absorption lines at $z = 0.3485 \pm 0.0005$ (see Figure \ref{fig_spec3}, left panel, fourth row). Given this result, we examined the plot of the spectrum of the source as reported by \cite{2021ApJS..254...26R} to understand the discrepancy. After careful examination, we are able to identify H$\alpha$ and the Calcium triplet at redshift zero, thus identifying the object as a star. This may be the result of an inaccurate pointing.

\subsection{NVSS J192502\texorpdfstring{$+$}{+}28154}
In Paper~III, we presented the spectra of NVSS J192502$+$28154 obtained with Lick/KAST during observations conducted on 9 June 2021, and 2 September  2021, with total exposure times of 4200 s and 7200 s, respectively, resulting in low S/N of 13 and 33. These spectra did not display any significant features. In the current work, we observed the source using Keck/ESI on 15 October 2021, achieving a significantly improved S/N of 65 with a total exposure time of 6300 s. Our observations reveal the presence of Mgb and NaID absorption lines, detected at a redshift of $z = 0.1636 \pm 0.0003$. The spectrum is shown in the right panel of the fourth row of Figure \ref{fig_spec3}.  

\subsection{1RXS J194246.3\texorpdfstring{$+$}{+}103339}
\begin{figure*}
   \centering
 \includegraphics[width=6.8truecm,height=5.75truecm]{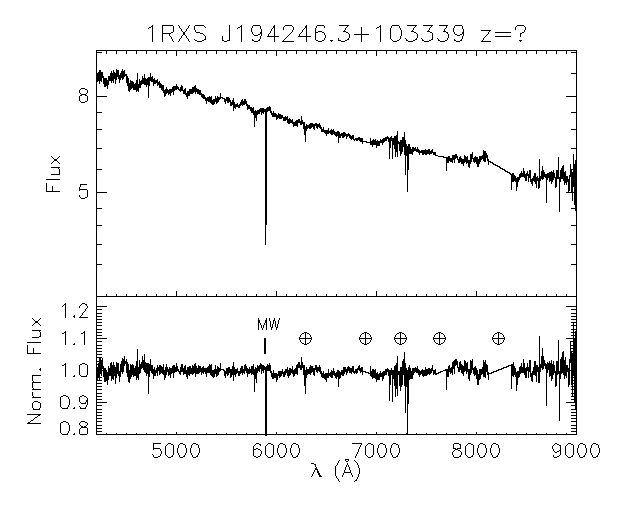}
 \includegraphics[width=6.8truecm,height=5.75truecm]{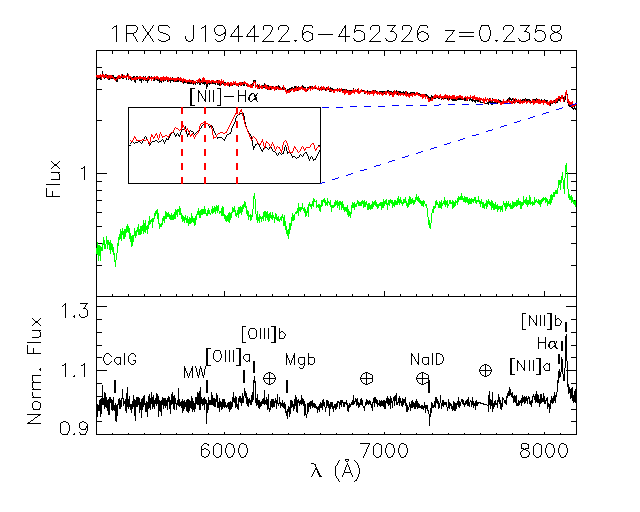}
 \includegraphics[width=6.8truecm,height=5.75truecm]{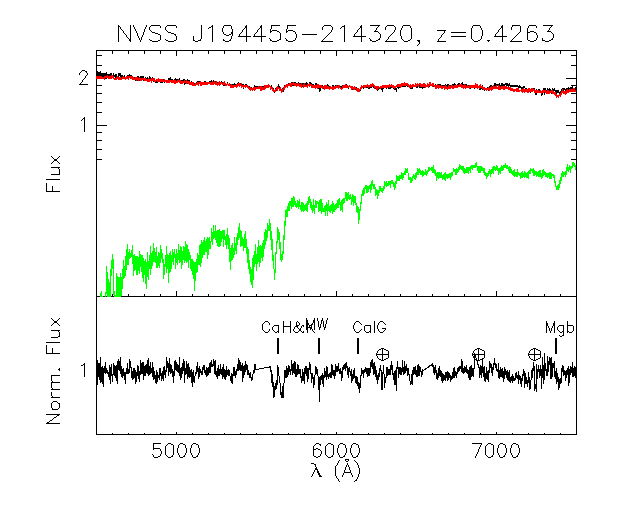}
 \includegraphics[width=6.8truecm,height=5.75truecm]{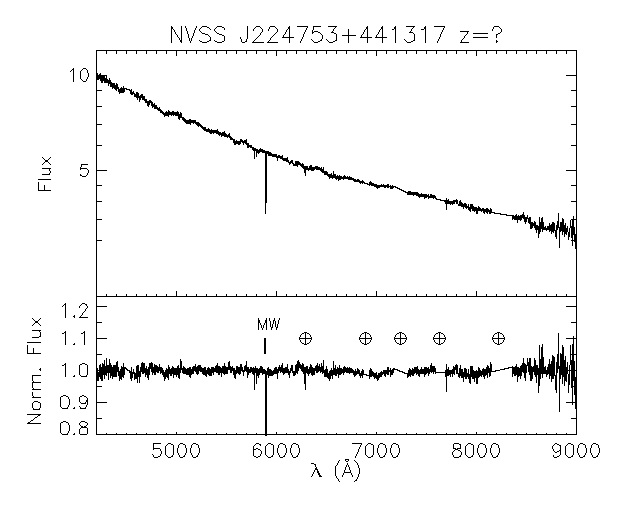}
 \includegraphics[width=6.8truecm,height=5.75truecm]{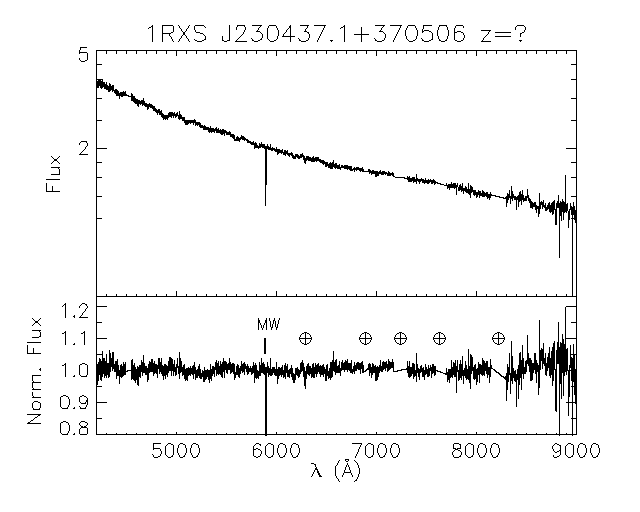}
\caption{Same as Fig. \ref{fig_spec1} for sources 25 to 29 in Table \ref{tabobs1}.}
\label{fig_spec4}
    \end{figure*}

\begin{figure*}[!h]
     \centering
  \includegraphics[width=6.truecm,height=5.75truecm]{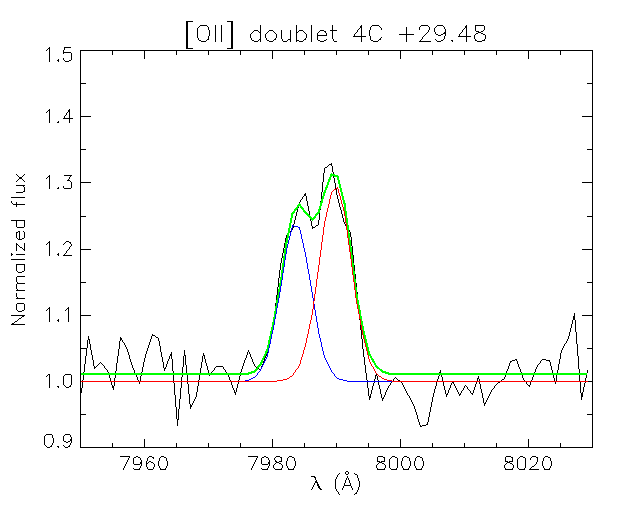}
  \includegraphics[width=6.truecm,height=5.75truecm]{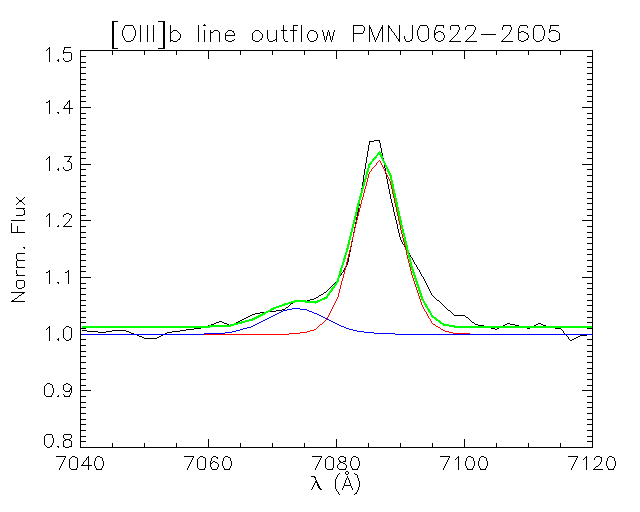}
  \includegraphics[width=6.truecm,height=5.75truecm]{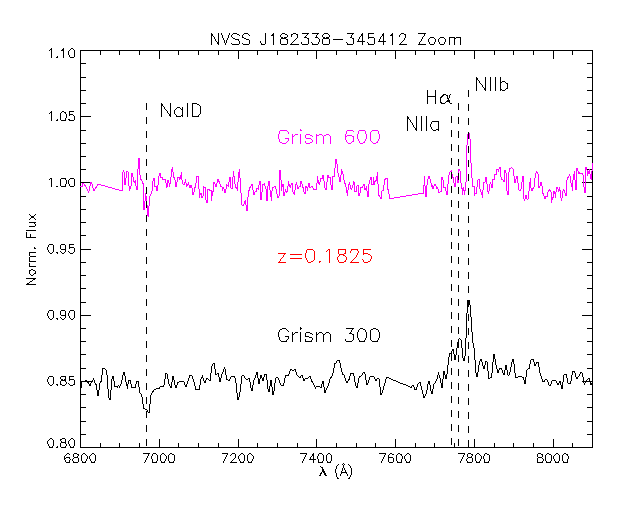}
\caption{ {\sl Left Panel}: Zoom in on the [OII] $\lambda$ 3727 feature detected in 4C +29.48 with Keck/ESI. As a single Gaussian cannot fit this feature, we fitted the two components of the doublets: $\lambda$ 3726 (blue line) and $\lambda$ 3729 (red line), with two gaussians. The sum of the two gaussians is shown by the green line. The intensity ratio of the two lines, F($\lambda$ 3729)/F($\lambda$ 3726), is $\sim$ 1.3 which is consistent with a thin plasma (see Section \ref{Dis_apen1}). {\sl Center Panel}: Zoom in on the [OIII] $\lambda$ 5007 feature detected in PMN J0622-2605 with VLT/FORS2. A single Gaussian cannot fit this feature, and there is a clear flux excess on the blue side as is frequently seen in this line. The excess is typically attributed to gas outflow \citep[see, e.g.][]{Sin22}. We thus fit two Gaussians to the profile, obtaining an acceptable $\chi^2/dof$ of 1.8. The stronger Gaussian is at the redshift of the system, while the weaker one is offset by -540 $\pm$ 40 km/s. The widths of the two Gaussians are 360 $\pm$ 10 km/s and 460 $\pm$ 50 km/s, respectively (see Section \ref{Dis_apen2}). {\sl Right Panel}: Normalized spectra (offset for clarity) of NVSS J182338-345412 obtained with VLT/FORS2 using Grism 600 (upper, magenta line) and Grism 300 (lower, black line). Overplotted are the features clearly identified in the spectra. The spectra are consistent. In the Grism 300 spectrum [NII] $\lambda$ 6548 and H$\alpha$ are detected additionally, suggesting slight variability in the non-thermal component between the two observations.}
          \label{fig_app_lines}
   \end{figure*}

A moderate S/N (45) observation of this source was conducted using VLT/FORS1 by \cite{2005A&A...438..949T}, revealing a featureless spectrum. On 4 June 2022, we observed this source with Keck/ESI for a total exposure time of 7200 s, obtaining a high S/N ($\sim$ 97). Our resulting spectrum is featureless, with no significant spectral lines detected and the redshift is still unknown (see Figure \ref{fig_spec4}, top left panel).

\subsection{1RXS J194422.6\texorpdfstring{$-$}{-}452326}
\citet{3HSP} estimated the photometric redshift of 1RXS J194422.6$-$452326 to be $z = 0.21$. Subsequent observations of this source by \citet{2024MNRAS.532.4785M}, using the South African Astronomical Observatory (SAAO) 1.9-m telescope, determined a spectroscopic redshift of $z = 0.236 \pm 0.002$, based on the identification of [OII] emission as well as Ca H\&K absorption lines. We conducted our observations of this source on 31 March 2022 using VLT/FORS, achieving a higher S/N of 75 over a total integration time of 500 s. Our high S/N spectrum revealed additional spectral features, including CaIG, Mgb, and NaID absorption lines, along with [OIII]a, [OIII]b, [NII]a, $H_\alpha$, and [NII]b emission lines. This analysis confirmed the redshift of the source as $z = 0.2358 \pm 0.0002$. The spectrum is presented in Figure \ref{fig_spec4} (upper right panel).

\subsection{NVSS J194455\texorpdfstring{$-$}{-}214320}
\cite{2021ApJS..254...26R} studied NVSS J194455$-$214320 with the Blanco 4-m Telescope, analyzing its optical spectrum and placing a lower limit on its redshift of $z > 0.41$ based on the detection of MgI and CaFe absorption lines. Subsequently, \cite{2023AJ....165..127G} observed this source with the SOAR telescope to measure the redshift, determining it to be $z = 0.426 \pm 0.001$. Our observations of this source were conducted with SALT/RSS on 5 June 2021, 10 July 2021, and 12 July 2021, with an exposure time of 2250 s per session, achieving a high S/N of 76. From our spectra, we identified CaHK, CaIG, and Mgb absorption lines at a redshift of $z = 0.4263 \pm 0.0003$. The spectrum is shown in Figure \ref{fig_spec4} (left panel, second row).

\subsection{NVSS J224753\texorpdfstring{$+$}{+}441317}
Featureless, low S/N spectra for this source were previously reported by \cite{2013ApJ...764..135S} and \cite{2015A&A...575A.124M}. In Paper~II, we conducted observations of this source using the Lick/KAST spectrograph, accumulating a total exposure time of 7200 s, but the resulting spectrum remained featureless. An imaging redshift estimate of $z = 0.34 \pm 0.07$, derived from observations with the Nordic Optical Telescope (NOT), was proposed by \cite{2024A&A...691A.154N}. We observed NVSS J224753$+$441317 on 15 October 2021 with Keck/ESI for a total exposure time of 6300 s and obtained a high S/N of 133. However, our spectrum lacked significant features (see Figure \ref{fig_spec4}, second row, right panel). We performed simple simulations which show that this non-detection is compatible with the above imaging redshift estimate provided that the R$_c$ absolute magnitude of the galaxy is fainter than -23.0.

\subsection{1RXS J230437.1\texorpdfstring{$+$}{+}370506}
A low S/N, featureless spectrum was previously obtained and analyzed by \cite{2013ApJ...764..135S}. In Paper~II, we reported two spectra of 1RXS J230437.1$+$370506, acquired with Lick/KAST on 30 August 2019 and 21 July 2020, with S/N values of 33 and 16, respectively, both of which were featureless. In this work, we observed its spectrum using Keck/ESI on 15 October 2021, achieving a significantly improved S/N with a total exposure time of 6300 s. The resulting high S/N ($\sim$ 95) spectrum and the redshift remains unknown, shown in the bottom panel of Figure \ref{fig_spec4}, remains featureless.

\section{Comparison with ZTF and ASAS-SN light curves}\label{lcurves}
Spectroscopic observations made during the photometric minimum can benefit from the reduced non-thermal emission from the jet at this time, enabling the host galaxy features to become more visible, which are then used to determine the redshifts \citep[see e.g.][]{2021MNRAS.504.5258B,2021A&A...650A.106G,2024A&A...683A.222D}.  We obtained the long-term light curves from the ZTF \citep{Masc19}\footnote{\url{http://www.ztf.caltech.edu}} and the All-Sky Automated Survey for SuperNovae (ASAS-SN) \citep{2014ApJ...788...48S,2017PASP..129j4502K}\footnote{\url{https://asas-sn.osu.edu//}}. Our goal was to determine whether our spectroscopic measurements were conducted during high, intermediate, or low activity states and to assess the impact of activity levels on redshift measurement efficiency.

We briefly summarize the characteristics of these two surveys. The ZTF, which uses the 48-inch Oschin Schmidt telescope at Palomar Observatory, achieves a sensitivity of r$\sim$20.6 (5$\sigma$ in 30 seconds) with a pixel scale of 1\arcsec, although its coverage is primarily limited to the northern hemisphere. In contrast, ASAS-SN reaches a sensitivity of g$\sim$18 (5$\sigma$ in 5 minutes) with a pixel scale of 8\arcsec and provides full-sky coverage through its global network. When ZTF measurements are available, they are preferred due to their lower uncertainties. If ZTF light curves are not available, we checked for the presence of ASAS-SN light curves. 

 The ZTF light curves for NVSS J105534$-$012617, 4C +29.48 and NVSS J184919$-$164723 have been rejected from our analysis. NVSS J105534$-$012617 is discarded due to an insufficient number of data points and the absence of nearby observations within 100 days. 4C +29.48 is omitted due to poor data quality, primarily due to its weakness. NVSS J184919$-$164723 is rejected due to contamination by a neighboring star. The ASAS-SN light curves for SUMSS J052542$-$601341, NVSS J080405$-$362919, NVSS J084121$-$355506, NVSS J105534$-$012617, IRXS J130737.8$-$425940, PMN\, J1544$-$6641, NVSS J163750$-$344915 and NVSS J182338$-$345412 have been excluded due to unreliability caused by contamination from nearby sources. Additionally, the light curves for NVSS J184919$-$164723 and IRXS J194422.6$-$452326 are not available in ASAS-SN. Tables \ref{tabLC} and \ref{tabLC2} show the time intervals between the spectroscopic measurements and the closest observational data collected by ZTF and ASAS-SN, respectively. These tables also provide the observed magnitudes, along with the median, maximum, and minimum magnitudes calculated over the entire period covered by ZTF and ASAS-SN, as well as the redshift or its lower limit determined in this study.

In our comparison with ZTF light curves, 13 out of 16 objects have been observed at magnitudes comparable to or greater than the median magnitude (indicating a fainter or low state). Among these 13 blazars, we have determined the redshifts for 4 sources through spectroscopic observations. The remaining 3 sources have been observed at magnitudes lower than the median (indicating a brighter or high state), with the redshift of one source successfully determined.

In comparison with ASAS-SN light curves, one out of two sources has been observed in a low state, while the other has been observed in a high state. We have measured the redshifts for both sources. 

\begin{table*}
\caption{\label{tabLC} Comparison with ZTF light curves in the r band: Note that 1RXS\,J094709.2$-$254056, 1RXS\,J171405.2$-$202747, and NVSS\,J194455$-$214320 each have three values corresponding to three separate observations (see Table \ref{tabobs1}). The columns include the following: (1) Source Name; (2) Delta time (days), representing the time difference between our observation date(s) and the closest ZTF observation; (3) magnitude at the nearest date; the source’s (4) median, (5) brightest, and (6) faintest magnitudes; and (7) the measured redshift.}
\centering
\begin{tabular}{lcccccc}
\hline\hline
Source Name  & Delta time & Closest  & Median   & Brightest & Faintest  & Redshift \\
              &  (days)   &  Mag.     &  Mag.   & Mag. &  Mag.  &   \\  
   (1)  & (2) & (3)    &  (4)  &  (5)   &  (6)  & (7) \\      
\hline 
1RXS\,J032342.6$-$011131$^*$ & 29.04  & 17.4  & 17.4  & 17.3  & 17.9 &  --- \\
RX\,J0338.4+1302$^*$ & 0.11  & 18.6  & 18.2  & 17.1  & 18.9 &  --- \\
GB6\,J0540+5823$^*$  & 0.87  & 17.9  & 17.5  & 16.8 & 18.4  &  --- \\
TXS\,0700$-$197 & 6.19 & 18.0 & 18.3 & 16.7 & 19.2 &  --- \\
RX\,J0819.2$-$0756$^*$ & 4.07 & 18.8 & 18.5 & 18.3 & 18.9 & 0.3241 \\
1RXS\,J094709.2$-$254056$^*$ & 1.28/0.27/6.61 & 16.6/16.6/16.8 & 16.7 & 15.9 & 17.1 & ---\\
NVSS\,J123341$-$014426 & 0.15 & 18.2 & 18.6 & 17.2 & 20.3 & ---\\
NVSS\,J160756$-$203942 & 8.09 & 18.3 & 18.5 & 17.7 & 19.7 & 0.5479\\
NVSS\,J164011+062827$^*$ & 4.94 & 18.0 & 18.1 & 17.8 & 18.4 & ---\\
1RXS\,J171405.2$-$202747$^*$ & 6.03/1.91/0.87 & 18.6/18.6/18.5 & 18.5 & 18.2 & 18.8 & 0.5222 \\
RX\,J1754.1+3212$^*$ & 1.16 & 16.9 & 16.6 & 15.9 & 17.3 & ---\\
NVSS\,J192502+28154$^*$ & 18.0 & 17.7 & 17.8 & 17.2 & 19.6 & 0.1636 \\
1RXS\,J194246.3+103339$^*$ & 0.88 & 17.3 & 16.5 & 15.9 & 17.4 & --- \\
NVSS\,J194455$-$214320$^*$ & 1.50/0.50/0.72 & 18.2/18.2/18.2 & 18.3 & 17.3 & 18.6 & 0.4263 \\
NVSS\,J224753+441317$^*$ & 0.02 & 17.3 & 17.3 & 16.3 & 17.8 & --- \\
1RXS\,J230437.1+370506$^*$ & 0.03 & 18.2 & 17.9 & 17.1 & 18.3 & --- \\
\hline
\end{tabular}
\tablefoot{The sources marked with an asterisk are those observed at magnitudes comparable to or greater than the median magnitude.}
\end{table*}

\begin{table*}
\caption{\label{tabLC2} Comparison with ASAS-SN light curves in the g band. Columns are the same as in Table \ref{tabLC}.}
\centering
\begin{tabular}{lcccccc}
\hline\hline
Source Name  & Delta time & Closest & Median   & Brightest & Faintest & Redshift \\
            &  (days)   &  Mag. & Mag. & Mag.  & Mag.  &    \\  
   (1)  & (2) & (3)    &  (4)  &  (5)   &  (6)  & (7) \\      
\hline 
PMN\, J0622-2605$^*$ & 3.06 & 17.2 & 17.0 & 15.7 & 17.5 & 0.4150 \\
1RXS\,J144037.4$-$384658 & 0.56 & 16.8 & 17.2 & 15.1 & 17.8 & 0.3583 \\
\hline
\end{tabular}
\end{table*}

\begin{table*}
\caption{\label{resultsPapers} The observed source counts, redshift values, and lower limit measurements (with uncertain results indicated in parentheses) are provided for different source groups, encompassing the observations reported in Paper~I, Paper~II, Paper~III, \citet{Gol24} and the current study (Paper IV). Note: the combined results take into account repeated observations (see discussion for details).}

\centering
\begin{tabular}{llllllll}
\hline\hline

Paper & Number of & Redshifts & Redshift & $z_{med}$ & Efficiency \\
& targets & ($z$) & lower limits & (with limits)   & \\
(1)  & (2) & (3)    &  (4)  &   (5)   &  (6)  \\  

\hline

I  & 19(0) & 11 (+1)$^{C}$ & 2(+1)$^{C}$  & 0.21(0.23)  & 11/19 ~~~~~~~~| 58\%\\
II & 25(8) & 14 (+1)$^{C1}$ & 2 &  0.37(0.38)  & 14/25 (33) | 56\% (42\%) \\
III & 24(17) & 12 (+1) & 2 &  0.39(0.42)  &  12/24 (41) | 50\% (29\%)\\
IV & 29(0) & 12 & 2 & 0.36(0.42) & 12/29 ~~~~~~~~| 41\% \\
\hline
\hline
Combined & 90(25) & 49(+1) & 8 &  0.32/0.36 & 49/90 ~~~~~~~| 54\% \\
\hline
\end{tabular}
\begin{minipage}{18 cm}
\small Notes: The columns are as follows: (1) Paper number; (2) Number of observed targets (targets observed with the Lick telescope) ; (3) Number of measured redshifts; (4) Lower limits on redshift; (5) Median redshift; and (6) Efficiency of redshift detection. Additionally, the combined results of the program, including repeated observations, are presented. The "+1" notation in columns 3 and 4 indicates the inclusion of either additional tentative redshifts or tentative lower limits in the overall count. C \& C1: These uncertain results are confirmed in Papers~III and Paper IV, respectively.
\end{minipage}
\end{table*}    

\section{Conclusions}\label{Conc}
Measurement of the redshifts of a significant number of {\em Fermi}-LAT-detected blazars with a hard LAT
spectrum is crucial, as they are strong candidates for CTAO observations. In this study, 29 BL Lac objects from the 3FHL catalog, identified as promising targets for future CTAO observations, were observed using the VLT, Keck II, and SALT telescopes. Among these 29 sources, we determined spectroscopic redshifts for 12 sources with values ranging from 0.1635 to 1.1427, which is the highest spectroscopically determined redshift in the sample. Furthermore, we identified 2 lower limits for redshifts: $z > 1.0196$ and $z > 1.4454$. Examining the quality of the collected spectra, 9 out of 29 BL Lacs have a high S/N. However, redshift measurements were obtained for only 1 of these 9 objects. For the remaining 20 sources, the S/N values ranged from 3 to 100, categorized as low or moderate, from which 11 firm redshifts and 2 lower limits were derived.

As previously noted in Papers II and III, achieving a high S/N is essential but not always sufficient for determining redshifts. The level of non-thermal jet emission often plays an even more critical role in feature detection. To assess the activity levels of the AGN in our sample, we collected ZTF and ASAS-SN light curves corresponding to the timing of our spectroscopic observations. Considering the 18 sources with reliable photometry near the date of spectroscopy (see Tables \ref{tabLC} and \ref{tabLC2}), 14 were in low state and 5 redshifts were measured while 4 were in high state and 1 redshift was measured. In this sample there is thus an indication that redshift determination efficiency is better in low state. Moreover, if we add to these our previous results in paper III, we have measured 12/26 redshifts in ‘low state’ and 2/11 in high state. Overall our results suggest that observations in the low state tend to facilitate redshift determination.


The median redshift from the new observations $z_{\rm\,med}^{\text{IV}}$ = 0.36 is comparable to the value reported in Paper~II ($z_{\rm\,med}^{\text{II}}$ = 0.37) and Paper~III ($z_{\rm\,med}^{\text{III}}$ = 0.39), but higher than the value in Paper~I ($z_{\rm\,med}^{\text{I}}$ = 0.21). We also calculated the combined results of the program to date, taking into account repeated observations (see Table \ref{resultsPapers}). In this study, we re-observed RX\,J0338.4+1302, RX\,J0819.2\texorpdfstring{$-$}{-}0756 (confirming its tentative redshift from Paper~II), and NVSS\,J163750\texorpdfstring{$-$}{-}344915. As a result, only one tentative redshift remains (B2\,0557+38 from Paper~III), and no tentative lower limits remain. To date, 90 independent targets have been observed, achieving 49 firm redshifts with an efficiency of approximately 54\% and a median redshift of $z_{med}^{\text{tot}}$ = 0.32. 

We have compared our redshift measurements with the upper limit redshift estimates derived using the EBL-attenuation method by \cite{2024MNRAS.527.4763D}. Our analysis indicated that the redshifts of three sources-NVSS J182338$-$345412, NVSS J192502+28154 and 1RXS J194422.6$-$452326 are broadly consistent with their respective upper limits of $0.29^{+0.12}_{-0.12}$, $0.39^{+0.38}_{-0.28}$ and $0.61^{+0.32}_{-0.27}$. In this study, we have identified a firm redshift of $z > 0.3$ for 8 sources. Previously, Paper~I reported 3 sources (2 firm redshifts and 1 tentative redshift), Paper~II reported 9 sources (8 firm and 1 tentative redshift), and Paper~III reported 10 sources (9 firm and 1 tentative redshift). Overall, 27 firm redshifts and 1 tentative redshift at $z > 0.3$ have been determined, with the tentative redshifts from Paper~I and Paper~II now confirmed in Paper~III and this study, respectively. These sources represent a valuable set of potential TeV emitters at $z > 0.3$, suitable for observations by VHE ground-based facilities like CTAO. They play a crucial role in understanding the luminosity function of BL Lac objects and in EBL studies. As such, this redshift measurement campaign is essential for supporting future CTAO observations of AGNs \citep[e.g.][]{CTAcons20,CTACons21-prop}.
%
%

\begin{acknowledgements}
 This research utilized the CTA instrument response functions provided by the CTA Consortium and Observatory. For further details, please refer to \url{https://doi.org/10.5281/zenodo.5163272} (version prod3b-v2). We sincerely acknowledge the financial support from the agencies and organizations acknowledged here: \url{http://www.cta-observatory.org/consortium_acknowledgments}, with special thanks to the U.S. National Science Foundation Grant PHY-2011420. The authors acknowledge and respect the deep cultural significance and reverence that the summit of Maunakea has always had within the indigenous Hawaiian community. We appreciate the privilege of conducting observations from this mountain. Some of the observations presented in this paper were obtained using the Southern African Large Telescope (SALT). This research has made use of the SIMBAD database, operated at CDS, Strasbourg, France. B.R. gratefully acknowledges support from the ANID BASAL project FB210003 and FONDECYT Postdoctorado 3230631. W.M. also acknowledges support from the ANID BASAL project FB210003. U. B. acknowledges a CNPq Productivity Grant nr. 309053/2022-6 and a FAPERJ Cientista Nosso Estado Grant nr. E-26/200.532/2023. The work of B.H. was supported by The National Research Foundation of Ukraine (project 2023.03/0149), Ukraine. 

\end{acknowledgements}

\bibliographystyle{aa} 
\bibliography{redshift}        

\begin{appendix}
\onecolumn
\section{}

\end{appendix}

\end{document}